\documentclass[useAMS,usenatbib]{mn2e}
\pdfoutput=1
\usepackage{graphicx}
\usepackage{natbib}
\bibpunct{(}{)}{;}{a}{}{,}
\usepackage{aas_macros}
\usepackage{amsmath}

\title[Multi-wavelength modelling of IRAS\,20126+4104]{The standard model of low-mass star formation applied to massive stars: multi-wavelength modelling of IRAS\,20126+4104}
\author[K. G. Johnston, E. Keto, T. P. Robitaille and K. Wood]{Katharine G.
Johnston$^{1, 2, 3}$\thanks{E-mail: johnston@mpia.de (KJ); \hfill\break
keto@cfa.harvard.edu (EK)}, Eric Keto$^{1}$, Thomas P. Robitaille$^{1}$ and Kenneth Wood$^{2}$
\\
$^{1}$Harvard-Smithsonian Center for Astrophysics, 60 Garden St, Cambridge, MA 02138, USA \\
$^{2}$School of Physics \& Astronomy, University of St Andrews, North Haugh, St Andrews, KY16 9SS, UK \\ 
$^{3}$Max-Planck-Institute for Astronomy, K\"onigstuhl 17, D-69117 Heidelberg, Germany \\}
\begin{document}

\date{}


\maketitle

\begin{abstract}
In order to investigate whether massive stars form similarly to their low-mass counterparts, we have used the standard envelope plus disc geometry successfully applied to low-mass protostars to model the near-IR to sub-millimetre SED and several mid-IR images of the embedded massive star IRAS\,20126+4104. We have used a Monte Carlo radiative transfer dust code to model the continuum absorption, emission and scattering through two azimuthally symmetric dust geometries, the first consisting of a rotationally flattened envelope with outflow cavities, and the second which also includes a flared accretion disc. Our results show that the envelope plus disc model reproduces the observed SED and images more accurately than the model  without a disc, although the latter model more closely reproduces the morphology of the mid-IR emission within a radius of 1.1" or $\sim$1800\,au. We have put forward several possible causes of this discontinuity, including inner truncation of the disc due to stellar irradiation, or precession of the outflow cavity. Our best fitting envelope plus disc model has a disc radius of 9200\,au. We find that it is unlikely that the outer regions of such a disc would be in hydrostatic or centrifugal equilibrium, however we calculate that the temperatures within the disc would keep it stable to fragmentation.

\end{abstract}


\section{Introduction}
Does the formation of a massive star differ significantly from that of a low mass star? This is the question at the core of all studies of high mass star formation to date. The fact that massive stars can have Kelvin-Helmholtz time-scales shorter than their accretion time-scales \citep{shu87}, leads to the consequence that their main-sequence radiation can affect their own formation. Hence, several processes such as radiation pressure \citep[e.g.][]{yorke02} and ionization \citep[e.g.][]{keto02} may halt, decrease or alter accretion on to the star. 

With the hypothesis that these differences do not radically change the formation of massive stars, our main aim is to find observational evidence to suggest the contrary. In this paper, we aim to determine whether the standard envelope plus disc model used to reproduce the spectral energy distributions (SEDs) of low mass stars \citep[e.g.][]{adams87, kenyon93, whitney03, wood02a} is able to reproduce the SED and observed infrared images of the nearby forming early B-type star IRAS\,20126+4104. 

\citet[][henceforth KZ10]{keto10} have recently modelled the line profiles of several millimetre-wavelength transitions of NH$_3$, C$^{34}$S and CH$_3$CN toward IRAS\,20126+4104, fitting these data with two models, the first consisting of an azimuthally symmetric rotationally flattened envelope, and the second also including a flared Keplerian disc. They found that the addition of a disc was required to reproduce the velocity structure of the accreting gas within 0.128\,pc of the star, providing a better fit than the without-disc model. Additionally, as the model with a disc was able to adequately reproduce the data, their results showed no evidence that ionization or radiation pressure are profoundly altering the dynamics of the surrounding accretion flow. In this work, we carry out the same experiment, but instead model the SED and observed infrared images of IRAS\,20126+4104. 

In comparison to the molecular line data, the SED and infrared images of IRAS\,20126+4104 probe different aspects of the circumstellar material. The molecular line observations are able to measure the velocity of the gas at different temperatures, however their density determinations depend on abundances that are highly uncertain. In contrast, the SED and infrared images probe the dust out to the edge of the envelope, which can be used to infer the structure of the gas. Hence, comparing the results from both these datasets will allow us to obtain a more complete picture of the star formation processes occurring in IRAS\,20126+4104.

IRAS\,20126+4104 is a well studied example of a nearby forming massive star \citep[$L_{\rm bol}=1.3\times$10$^4$\,L$_{\odot}$, B0.5 star,][]{cesaroni99}. Found within the Cygnus-X star forming region, it is assumed to be at the same distance as the Cygnus OB associations \citep[1.7\,kpc,][]{massey91,wilking90}. Several observations of molecular tracers at millimetre wavelengths have uncovered a $\sim$0.25\,pc NW-SE inner jet feeding a large-scale North-South bipolar outflow \citep{cesaroni97,cesaroni99,su07,lebron06,shepherd00}. In addition, high resolution centimetre continuum images of the ionized gas toward IRAS\,20126+4104 have been presented by \citet{hofner07}.

Observations of transitions of C$^{34}$S, CH$_3$CN, CH$_3$OH and HCO$^+$ \citep{cesaroni97,cesaroni99,cesaroni05b}, and NH$_3$ \citep[][KZ10]{zhang98} have uncovered a rotating flattened disc-like structure, perpendicular to the inner flow, with a radius between 800 and 12,000\,au, and a gas mass estimated from both dust continuum or molecular line observations of 0.65-10\,M$_{\odot}$.
Further evidence of a disc is also provided by near and mid-infrared images of IRAS\,20126+4104 \citep{sridharan05,de-buizer07}, which show a dark lane with a similar size and orientation to the disc seen in radio observations. A large (radius$\sim$1000\,au) rotating structure traced by OH masers was also found by  \citet{edris05}, which is coincident with the dark lane observed in the infrared. However, recently \citet{de-wit09} have presented a high-resolution (0.6") 24.5$\mu$m image of IRAS\,20126+4104, at a slightly longer wavelength than those presented in \citeauthor{de-buizer07} (\citeyear{de-buizer07}, 12.5 and 18.3$\mu$m), which does not contain a dark lane. 

The infrared to sub-millimetre SED of IRAS\,20126+4104 has been previously modelled on several occasions. Both \citet{cesaroni99} and \citet{shepherd00} modelled the far-IR to millimetre SED using a simple greybody to determine the dust temperature and the dust emissivity index. \citet{williams05}, \citet{de-wit09} and \citet{van-der-tak00} have also modelled the SED of IRAS\,20126+4104 using spherically symmetric models, to determine the properties of the envelope. More recently, \citet{hofner07} modelled IRAS\,20126+4104 using the radiation transfer code previously employed in \citet{olmi03}. Their model consisted of a spherical halo with a power-law density between radii of 850\,au and 0.54\,pc, plus a constant density cylindrical edge-on disc with radii between 34-850\,au and a height of 640\,au. The model provided a good fit to the data down to $\sim$8$\mu$m, however, they suggested that adding further components to the model, such as bipolar cavities and a more complex disc geometry, would improve the fit, especially in the near- to mid-IR section of the SED. In this paper, we improve upon previous infrared modelling and derive complimentary information about the properties of IRAS\,20126+4104 not previously obtainable from molecular line observations.

We present the observed SED, infrared images and measured brightness profiles of IRAS\,20126+4104 in Section \ref{data}; Section \ref{dustcode} describes the radiation transfer dust code used to model them; Section \ref{modelling} details the modelling, including model input assumptions, fitting and the genetic search algorithm used to search for the best-fitting set of model parameters; and Section \ref{results} presents our results, including a comparison to the results of KZ10. We give our conclusions in Section~\ref{conclusions}.

\section{The Data}
\label{data}
\subsection{Near-IR to sub-millimetre SED}
\begin{figure*}
\includegraphics[width=5.in]{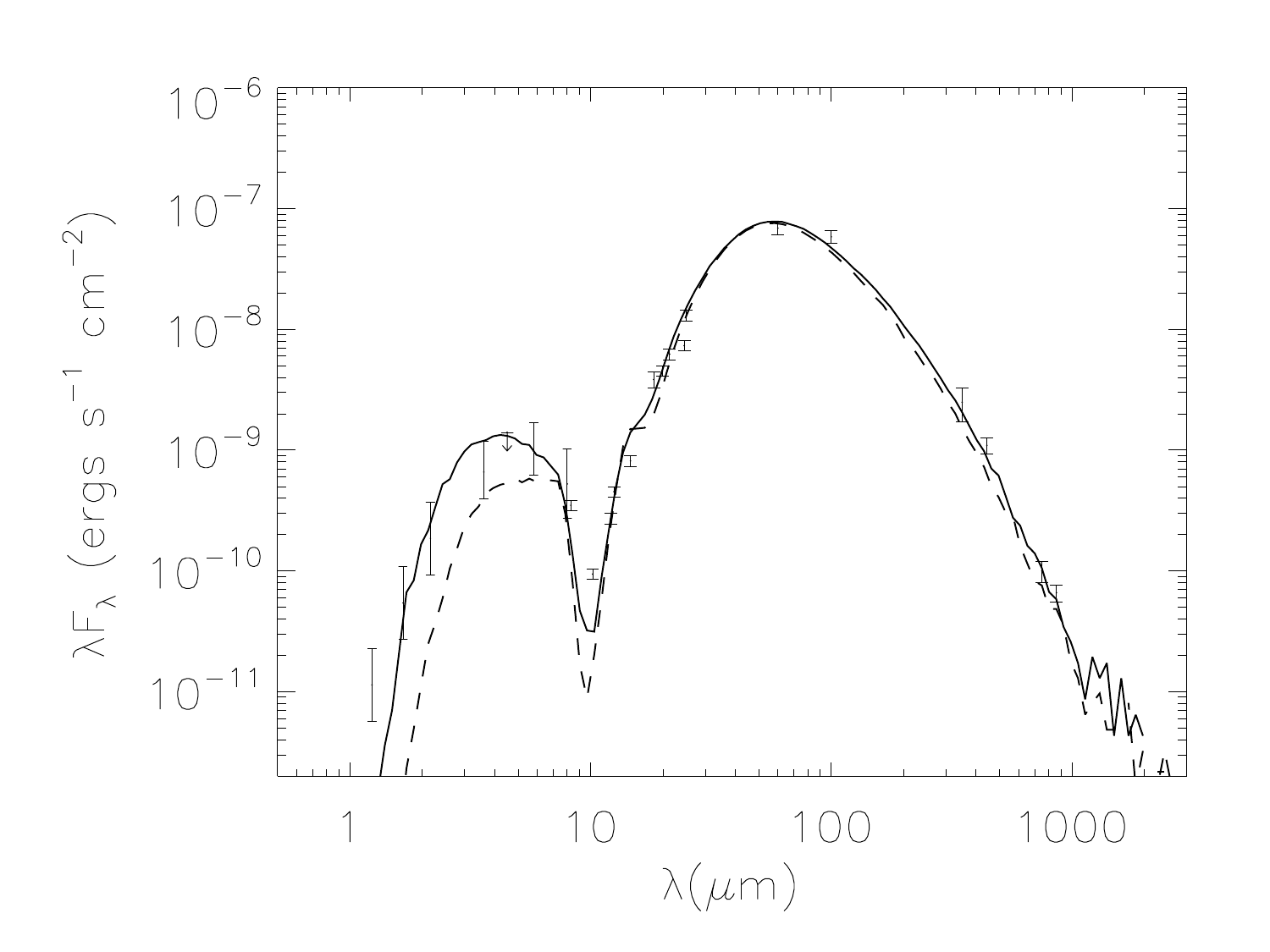}
 \caption{
The Spectral energy distribution (SED) of IRAS\,20126+4104. The best-fitting models for an envelope plus disc and without disc (overplotted solid and dashed lines respectively) are discussed in Section~\ref{results}. The errors shown are those reset to 10\% if the error in the measured flux was $<$10\%, and the errors in the 2MASS JHK fluxes to a factor of two. The IRAC 4.5$\mu$m flux was assumed to be an upper limit due to possible contamination by H$_2$ line emission.
}
 \label{seddata}
\end{figure*}

The observed near-IR to sub-millimetre SED of IRAS\,20126+4104 is shown in Figure \ref{seddata}. The overplotted solid and dashed lines (respectively the best-fitting models for an envelope plus disc and without a disc) are discussed further in Section~\ref{results}. The SED data points were collected from the literature; Table \ref{obstable} lists the wavelength, flux and reference for the data displayed in Figure~\ref{seddata}. If the fluxes were derived from apparent magnitudes, these are also given in Table \ref{obstable}. We attempted to make sure all fluxes contained all of the flux from the source. For instance, we used the 2 Micron All Sky Survey (2MASS) extended source catalog fluxes for the near-IR.

\begin{table*}
 \centering
 \begin{minipage}[t]{5.5truein}
  \caption{Observed near-IR to sub-millimetre fluxes for IRAS\,20126+4104, collected from the literature. Apparent magnitudes are given where the flux was derived from these magnitudes. References are given in parentheses: (1) \citet{jarrett00}, (2) this work, (3) \citet{qiu08}, (4) \citet{price01} , (5) \citet{cesaroni99}, (6) \citet{de-buizer07}, (7) \citet{de-wit09}, (8) \citet{joint-iras-science94} 
  }
  \begin{tabular}{@{}llllll@{}}
  \hline
  Wavelength & Flux & Apparent& Description / Origin \\
 ($\mu$m) & (mJy) & magnitude &  \\
 \hline
1.235 & 4.7 $\pm$ 49\% & 13.939 $\pm$ 0.527 & 2MASS Extended Source Catalog, J band (1)\\
1.662 & 30.1 $\pm$ 8.8\% & 11.333 $\pm$ 0.095 & 2MASS Extended Source Catalog, H band (1) \\
2.159 & 134 $\pm$ 2.3\% & 9.242 $\pm$ 0.025 & 2MASS Extended Source Catalog, K band (1) \\
3.6 & 790 &  & IRAC band 1, photometry: (2), image: (3) \\
4.5 & 2.1$\times10^{3}$ & & IRAC band 2, photometry: (2), image: (3) \\
5.8 & 1.9$\times10^{3}$ & & IRAC band 3, photometry: (2), image: (3) \\
8.0 & 1.4$\times10^{3}$ & & IRAC band 4, photometry: (2), image: (3) \\
8.28	 & 965 $\pm$ 4.2\% & & MSX band A (4) \\
10.2 & 320 & & UKIRT, MAX (5) \\
12.13 & 1.10$\times10^{3}$ $\pm$ 8.2\% & & MSX band C (4) \\
12.5 & 1.89$\times10^{3}$ $\pm$ 10\% & & Gemini North, Michelle (6) \\
14.65 & 3.98$\times10^{3}$ $\pm$ 6.2\% & & MSX band D (4) \\
18.3 & 2.35$\times10^{4}$ $\pm$ 15\% & & Gemini North, Michelle (6) \\ 
19.9 & 3.0$\times10^{4}$ & & UKIRT, MAX (5) \\
21.3 & 4.43$\times10^{4}$ $\pm$ 6.0\% & & MSX band E (4) \\ 
24.5 & 6.0$\times10^{4}$ & & Subaru telescope, COMICS (7) \\
25 & 1.09$\times10^{5}$ $\pm$ 5\% & & IRAS 25$\mu$m (8) \\
60 & 1.38$\times10^{6}$ $\pm$ 12\% & & IRAS 60$\mu$m (8)\\
100 & 1.95$\times10^{6}$ $\pm$ 13\% & & IRAS 100$\mu$m (8) \\
350 & 2.92$\times10^{5} \pm$ 9.0$\times10^{4}$ & & JCMT, SCUBA (5) \\
443 & 1.62$\times10^{5} \pm$ 2.4$\times10^{4}$ & & JCMT, SCUBA (5) \\
750 & 2.5$\times10^{4} \pm$ 5$\times10^{3}$& & JCMT, SCUBA (5) \\
863 & 1.9$\times10^{4} \pm$ 3$\times10^{3}$ & & JCMT, SCUBA (5) \\
\hline
\end{tabular}
\label{obstable}
\end{minipage}
\end{table*}

\begin{figure*}
\includegraphics[width=7.in]{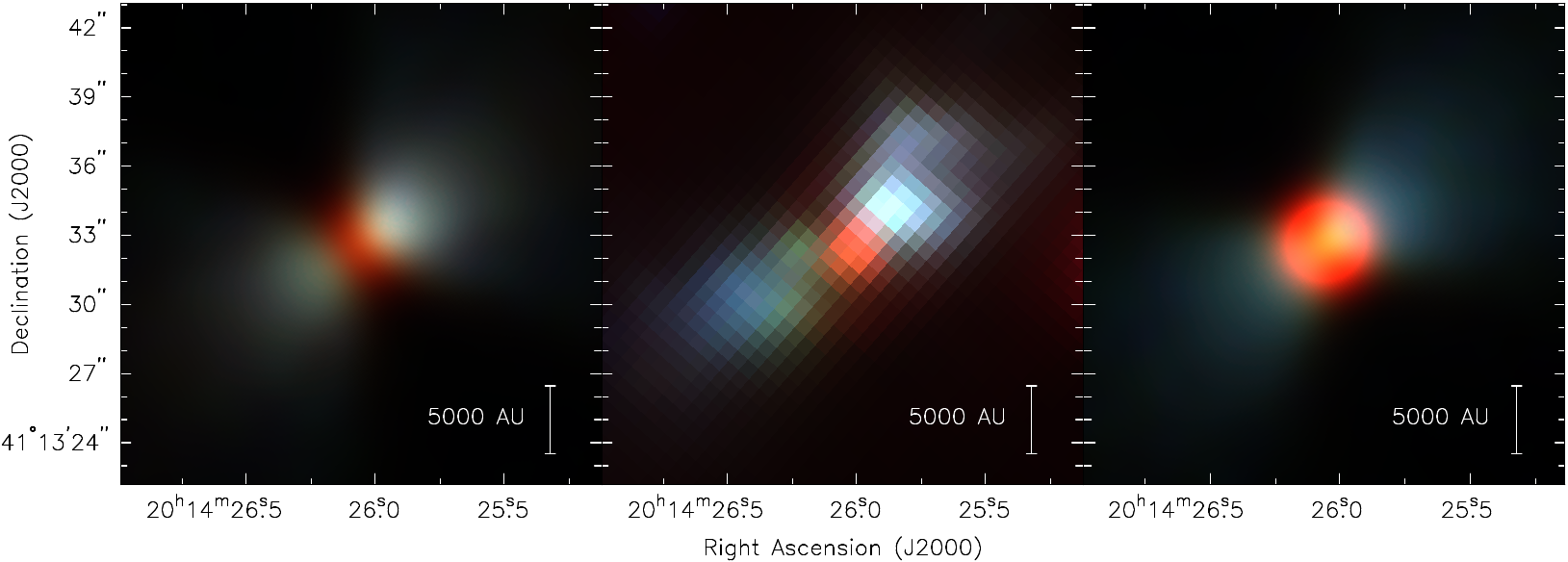}
 \caption{
\textit{Left panel:} Model IRAC image for envelope plus disc model. \textit{Middle panel:} Observed IRAC three-colour RGB image of IRAS\,20126+4104, previously presented by \citet{qiu08}. \textit{Right panel:} Model IRAC image for envelope without disc model. The model images have been normalised to the total integrated fluxes given in Table \ref{obstable}, so that the morphology of the emission can be easily compared. Stretch: red: 8\,$\mu$m, 0-1200 MJy\,sr$^{-1}$; green: 4.5\,$\mu$m, 0-1800 MJy\,sr$^{-1}$, blue: 3.6\,$\mu$m, 0-670 MJy\,sr$^{-1}$}
 \label{IRACrgb}
\end{figure*}

\subsection{Infrared images}
A three-colour Infrared Array Camera (IRAC) image of the mid-IR emission observed toward IRAS\,20126+4104, previously published by \citet{qiu08}, is shown in the middle panel of Figure \ref{IRACrgb}, where red: 8.0$\mu$m, green: 4.5$\mu$m and blue: 3.6$\mu$m. In this figure, two lobes of emission oriented similarly to the NW-SE jet observed by \citet{su07} can be seen, which are most obvious at 3.6$\mu$m and 4.5$\mu$m, while the 8.0$\mu$m IRAC emission mostly traces the central source. 

To obtain IRAC fluxes for IRAS\,20126+4104, we performed irregular aperture photometry on the images previously published in \citet{qiu08}, using the post-Basic Calibrated Data (post-BCD) mosaics produced by the \textit{Spitzer} Science Center (SSC) pipeline version S18.7. The photometry was carried out on the short exposure (0.4\,s) images, as there were several artifacts in the longer exposure (10.4\,s) mosaics. The apertures were drawn to best avoid any foreground stars in the images. 

Measurement uncertainties were $<$5\% at 3.6 and 4.5\,$\mu$m, $\sim$10\% at 5.8\,$\mu$m, and $\sim$30\% at 8\,$\mu$m, due to the increasing variance of the background in the longer wavelength bands. In addition, this should be combined with a calibration uncertainty of 10\% for pipeline post-BCD mosaics\footnote{Section 4.3 of the IRAC Instrument Handbook, version 1.0}. Further, due to scattering of flux throughout the image by the IRAC optics, fluxes of extended sources are highly uncertain. Therefore, at a conservative estimate, these fluxes are accurate to within a factor of two. However, this is adequate for the purpose of fitting a model to the observed SED.

\begin{figure*}
\includegraphics[width=6.in,angle=0]{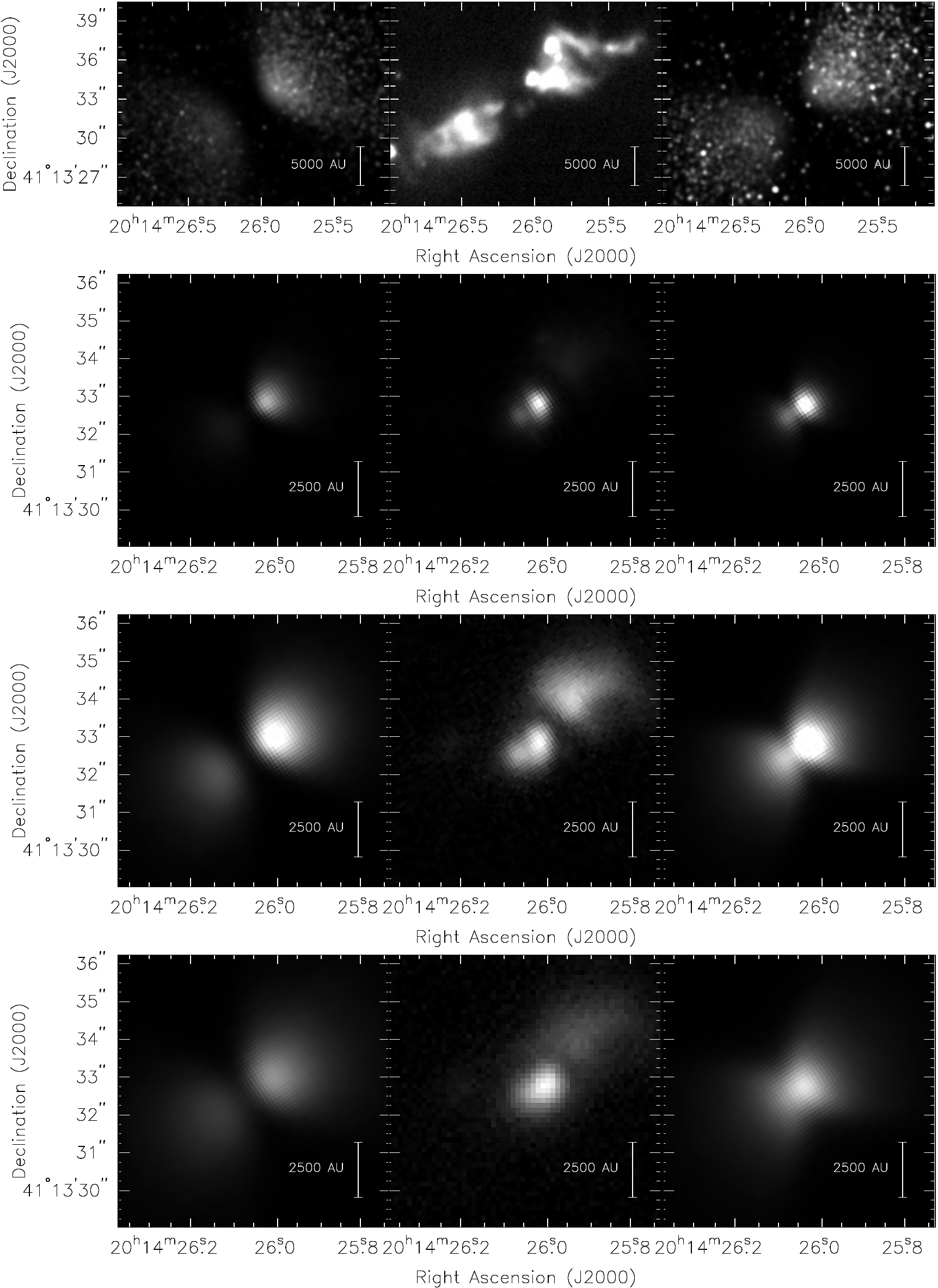}
 \caption{Observed (middle panels) and corresponding envelope plus disc and without disc model images (left and right panels respectively) for several near- and mid-IR observations of IRAS\,20126+4104. From top to bottom: K band \citep{sridharan05}, 12.5$\mu$m and 18.3$\mu$m \citep{de-buizer07}, and 24.5$\mu$m \citep{de-wit09}. The model images have been normalised to the total integrated fluxes given in Table \ref{obstable}, so that the morphology of the emission can be easily compared. Image stretches (top to bottom): 0-80.7 MJy\,sr$^{-1}$, 0-8.98 $\times 10^4$ MJy\,sr$^{-1}$, 0-2.67$\times 10^5$ MJy\,sr$^{-1}$, 0-2.03$\times 10^6$ M\,Jy\,sr$^{-1}$}
 \label{fig:combined}
\end{figure*}

The middle panels of Figure \ref{fig:combined} also show the observed K-band, 12.5\,$\mu$m, 18.3\,$\mu$m, and 24.5\,$\mu$m images presented in \citet{sridharan05}, \citet{de-buizer07} and \citet{de-wit09}, along with corresponding model images (left and right panels) which will be discussed in Section \ref{results}. 

\subsection{Brightness profiles}
\begin{figure*}
\includegraphics[width=5in,angle=0]{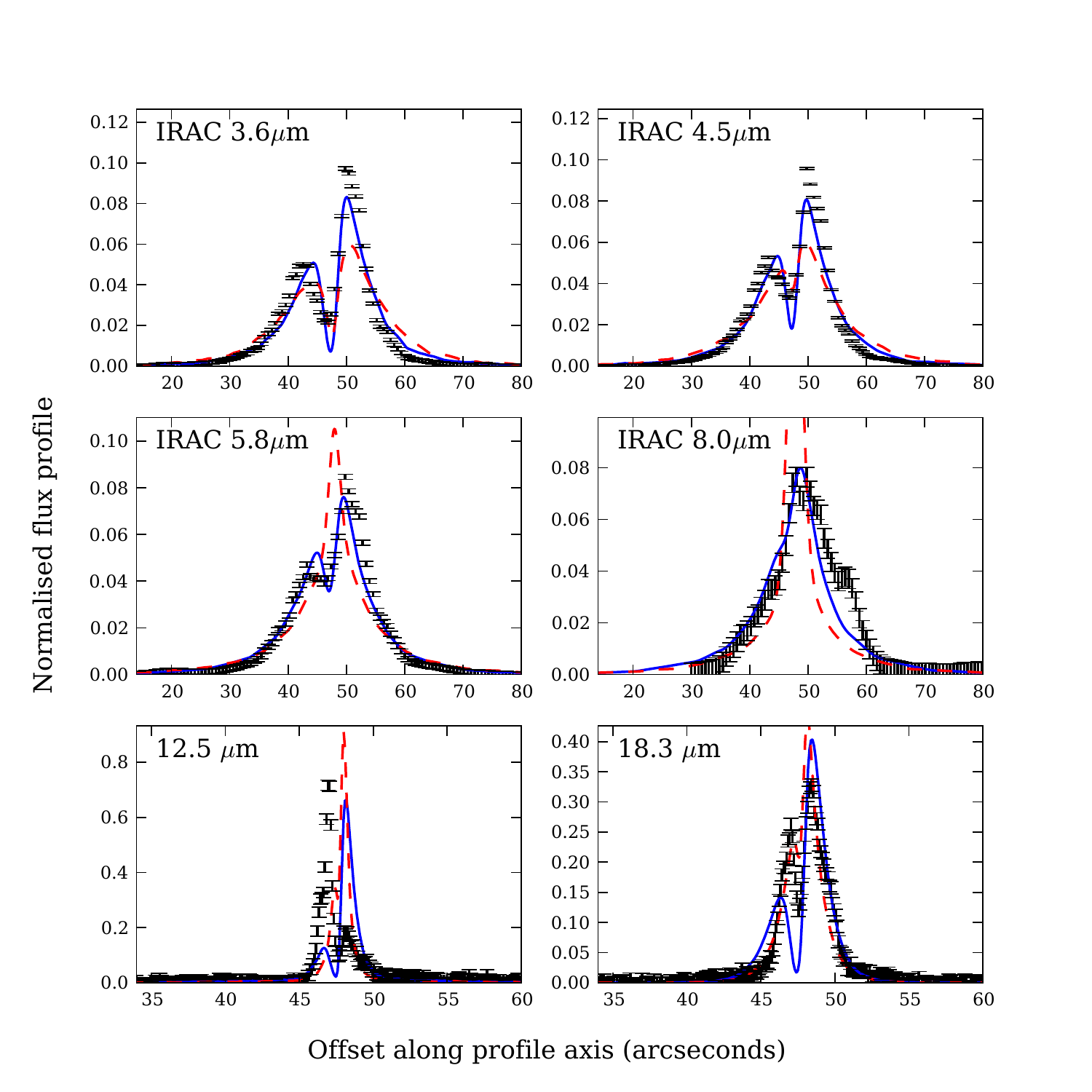}
 \caption{Black error bars: normalised flux profiles for the four IRAC bands and the two wavelengths observed by \citet{de-buizer07}: 12.5 and 18.3\,$\mu$m. Blue solid and red dashed lines: the profiles of the best-fitting models of both the observed SED and profiles, for envelope plus disc and without disc models respectively.}
 \label{fig:profiles}
\end{figure*}

Flux profiles of IRAS\,20126+4104, summed across a 15.1" thick strip aligned with the source outflow axis (PA=303.5 degrees), were measured for the four IRAC bands and at the two wavelengths observed by \citet{de-buizer07}: 12.5 and 18.3\,$\mu$m. The normalised profiles are shown in Figure~\ref{fig:profiles} along with the best-fitting models to both the SED and profiles, for envelope plus disc and without disc models (blue solid and red dashed lines respectively), which will be discussed further in Section \ref{results}. The errors in the profiles shown in each panel of Figure~\ref{fig:profiles} reflect the uncertainty in the background underlying the source emission due to background fluctuations, and are calculated as the standard deviation of the profiles measured in two strips either side of the main profile, which we assumed to contain minimal source flux. 

\section{The radiation transfer dust code}
\label{dustcode}

In this paper, we model the SED of IRAS\,20126+4104 between near-IR and sub-mm wavelengths using the 2D Monte Carlo dust radiation transfer code previously employed in \citet{wood02b} and modified in \citet{akeson05}. For a given source and surrounding dust geometry sampled in a spherical-polar grid, the code models the anisotropic scattering and thermal emission by/from dust, calculating radiative equilibrium dust temperatures \citep[using the technique of ][]{bjorkman01}, and producing SEDs and multi-wavelength images. Dust and gas are assumed to be coupled in the code, from which we aim to probe the bulk of the material which surrounds IRAS20126+4104. In this section, we first describe the density structure of the disc, envelope and outflow cavity in the model, then we detail the heating of the star via accretion on to its surface, and the dissipation of energy via disc accretion. 

We model the circumstellar geometry of IRAS\,20126+4104 with the same envelope plus disc geometry successfully employed to model the SEDs and scattered light images of low-mass protostars \citep[e.g.][]{robitaille07,wood02a,whitney03}, and also the molecular line emission from IRAS\,20126+4104 (KZ10). The two dimensional flared disc density is described between inner and outer radii $R_{\rm min}$ and $R_{\rm disc}$ by 
\begin{equation}
\rho_{\rm disc} (\varpi,z)=\rho_0 \left[ 1 - \sqrt{\frac{R_\star}{\varpi}}\right] \left (\frac{R_\star}{\varpi}\right )^{\alpha}
\exp{ \left\{ -{1\over 2} \left[\frac{z}{h( \varpi )}\right]^2 \right\}  }
\; ,
\label{discdensity}
\end{equation}
where $R_{\star}$ is the stellar radius, $\varpi$ is the cylindrical radius, $z$ is the height above the disc midplane and $\rho_0$ is set by the total disc mass $M_{\rm disc}$, by integrating the disc density over $z$, $\varpi$ and $\phi$, the azimuthal angle. The scaleheight $h(\varpi)$ increases with radius: $h(\varpi)=h_0\left ( {\varpi /{R_\star}} \right )^\beta$ where $h_0$ is the scaleheight at $R_\star$ and the disc is assumed to be in hydrostatic equilibrium at the dust sublimation radius. We have assumed the parameter $\beta=1.25$, derived for irradiated discs around low mass stars \citep{dalessio98}, and have taken $\alpha=2.25$, to provide a surface density of $\Sigma\sim \varpi^{-1}$.

The density of the circumstellar envelope is taken to be that of a rotationally flattened collapsing spherical cloud 
\citep[][]{ulrich76,terebey84} with radius $R^{\rm max}_{\rm env}$,
\begin{equation}
\begin{split}
\rho_{\rm env} (r,\mu) = {{\dot M_{\rm env}}\over {4\pi}\left({GM_\star R_c^3} \right)^{1/2}}
\left( {{r}\over{R_c}}\right)^{-3/2}\left(1+{{\mu}\over{\mu_0}} \right)^{-1/2} \\
\times \left( {{\mu}\over{\mu_0}}+{{2\mu_0^2R_c}\over{r}}\right)^{-1}
\end{split}
\end{equation}
where $\dot M_{\rm env}$ is the accretion rate through the envelope, $M_\star$ is the stellar mass, $r$ is the spherical radius, $R_c$ is the centrifugal radius, $\mu$ is the cosine of the polar angle ($\mu=\cos\theta$), and $\mu_0$ is the cosine of the polar angle of a streamline of infalling particles in the envelope as $r\rightarrow\infty$. The equation for the streamline is given by
\begin{equation}
\mu_0^3 + \mu_0(r/R_c-1)-\mu(r/R_c)=0\; 
\end{equation}
which can be solved for $\mu_0$. In our model, we assume the centrifugal radius is also the radius at which the disc forms $R_c = R_{\rm disc}$.

To reproduce the near-IR \citep{sridharan05} and IRAC \citep{qiu08} images of IRAS\,20126+4104, we also include a bipolar cavity in the model geometry with density $\rho_{\rm cav}$, and a shape described by
\begin{equation}
z(\varpi) = z_0 \varpi^{b_{\rm cav}}
\end{equation}
where the shape of the cavity is determined by the parameter $b_{\rm cav}$ which we set to be $b_{\rm cav}=1.5$, and $z_0$ is chosen so that the cavity half-opening angle at $z=10,000$\,au is $\theta_{\rm cav}$.

The inner radius of the dust disc and envelope $R_{\rm min}$ was set to the dust destruction radius $R_{\rm sub}$, found empirically by \citet{whitney04c} to be 
\begin{equation}
R_{\rm sub} = R_{\star}(T_{\rm sub}/T_{\star,\rm{acc}})^{-2.1}
\end{equation}
where the temperature at which dust sublimates is assumed to be $T_{\rm sub}=1600$K, and $T_{\star,\rm{acc}}$ is the effective temperature of the star, including heating by accretion shocks on to the stellar surface. $T_{\star,\rm{acc}}$ was found by assuming that the material accreting through the disc is in free-fall from the inner radius of the gas disc $R_{\rm gas}$ to $R_{\star}$ and that half of the energy lost by free-fall goes into heating the stellar surface \citep{calvet98}. The other half of this energy is assumed to go into X-ray emission, which is not included in the code. As virtually all of the X-rays are reprocessed for embedded sources like IRAS\,20126+4104, the determined disc accretion rates may be a factor of two too high, as without X-ray emission the disc accretion rate has to be a factor of two higher to reproduce the same accretion luminosity. We assume $R_{\rm gas}=5R_{\star}$, in keeping with findings for discs around low-mass stars \citep[][]{shu94}, however we note that magnetic disc truncation may not occur for massive stars. Therefore, the accretion luminosity which goes into heating the stellar surface is
\begin{equation}
L_{\rm heat} = \frac{GM_{\star}\dot{M}_{\rm disc}}{2} \left(\frac{1}{R_{\star}}- {\frac{1}{R_{\rm gas}}}\right) , 
\end{equation}
and the effective temperature of the star is then
\begin{equation}
T_{\star,\rm{acc}}=T_{\star}\left( \frac{L_{\star}+L_{\rm heat}}{L_{\star}} \right)^{1/4}
\,,
\end{equation}
where $L_{\star}$ is the luminosity of the star.

In addition to luminosity from the star, the dust disc emits due to accretion. The energy dissipated through dust radiation in a volume element between $R_{\rm min}$ and  $R_{\rm disc}$ is given by the $\alpha$-disc prescription \citep[e.g.][]{shakura73,kenyon87,whitney03},
\begin{equation}
\frac{dL_{\rm acc}}{dV} = \frac{3GM_{\star}\dot{M}_{\rm disc}}{\sqrt{32 \pi^3} \varpi^3 h(\varpi)} \left ( 1- \sqrt{\frac{R_{\star}}{\varpi}} \right ) \exp{ \left\{ -{1\over 2} \left [ \frac{z}{h( \varpi )} \right ]^2 \right\} },
\end{equation}

with the wavelength of this emission determined by the local dust temperature and opacity.

For the circumstellar dust properties we adopt the dust opacity and scattering properties determined by \citeauthor{kim94} (\citeyear{kim94}, hereafter KMH), which have a grain size distribution with an average particle size slightly larger than the diffuse ISM, and a ratio of total-to-selective extinction $R_{\rm{V}} = 3.6$.

In order to compare the models to existing observations, synthetic model images were produced by the dust code using the \textit{peel-off} approach \citep[originally described in][]{yusef-zadeh84}, whereby each time a photon interacts, the probability that the photon reaches the observer, and therefore the emergent intensity from that point, is found.

To produce the model IRAC images, we used the IRAC filter spectral response curves, obtained from the SSC website\footnote{http://ssc.spitzer.caltech.edu/irac/calibrationfiles/spectralresponse}, and convolved the images with the FWHM of the IRAC point-spread function (PSF) for each band. 

To produce the images at the four wavelengths shown in Figure \ref{fig:combined} (K band, 12.5$\mu$m, 18.3$\mu$m and 24.5$\mu$m), we applied the correct band pass filters, obtained from the United Kingdom infrared telescope (UKIRT)\footnote{http://irtfweb.ifa.hawaii.edu/$\sim$nsfcam/filters.html}, the Gemini telescope web pages\footnote{http://www.gemini.edu/sciops/instruments/michelle/imaging/filters}, and de Wit (private communication), and convolved the images to the observed resolutions (K band: 0.3", 12.5$\mu$m: 0.33", 18.3$\mu$m: 0.48" and 24.5$\mu$m: 0.6").

\section{Modelling}
\label{modelling}

In this section, we firstly describe the model input assumptions, including the selection of sensible physical parameter ranges for IRAS\,20126+4104. Secondly, we detail the $\chi^2$ fitting procedure carried out to compare the models to the data, and finally we outline the genetic search algorithm which was used to search parameter space for the most suitable model.

\subsection{Input assumptions: model parameters}
\label{input}

A set of plausible ranges for parameters describing the model, in which the genetic algorithm (described in Section~\ref{geneticshort}) searched for the best fit, is given in Table \ref{ranges}. Ranges for the first six parameters - stellar mass, envelope outer radius, cavity half-opening angle, inclination, disc mass and disc radius - were chosen using results from the literature, specifically \citet{cesaroni97,cesaroni99b,cesaroni05b}, \citet{de-buizer07}, \citet{edris05}, \citet{qiu08} and \citet{zhang98}. 
Large ranges were assumed for the disc mass and radius, as although a disc-like rotating structure has been detected at the centre of IRAS\,20126+4104 \citep[e.g.][]{cesaroni05b}, it is possible that these observations are also tracing part of the envelope.

The following three parameters in Table \ref{ranges} - the envelope and disc accretion rates and cavity ambient density - were also given large but physically plausible ranges.

The final two parameters, the stellar radius and temperature, were determined from the sampled stellar mass for that model. To do this, we used the $\dot{\rm{M}}\propto$~M$^2$ evolutionary track presented in Figure 1 of \citet{keto06}, which results from the stellar evolution models of \citet[][]{chieffi89}. These calculations show that, before taking into account heating from accretion shocks, the protostellar temperature and luminosity for stars which reach masses greater than 4M$_{\odot}$ are equivalent to a zero age main-sequence (ZAMS) star. This can be seen in Figure 1 of \citet{keto06}, where for all the evolutionary tracks (or accretion rates), stars which reach masses greater than 4M$_{\odot}$ follow the main-sequence line until their accretion is unable to counteract the production of helium in their cores, and they begin their post-main-sequence evolution. 

Once the stellar mass, radius and temperature were determined, we calculated the stellar temperature including heating from shocks $T_{\star,\rm{acc}}$ and the surface gravity of the star. We then used the nearest solar metallicity ($Z=0.02$) stellar model atmosphere from the grid of \citet{castelli04}, or a blackbody spectrum if the temperature was larger than 50,000\,K, to provide the spectrum of the central source. 

\subsection{Model fitting}
\label{fitting}

As part of the genetic algorithm described in Section~\ref{geneticshort}, the models were fit to the data using the fitting algorithm described in \citet{robitaille07}, where $ {A_{\rm v}}$, the external interstellar visual extinction, was a free parameter in the fit.
We set the distance to be 1.7\,kpc, in keeping with the adopted distance to IRAS\,20126+4104 \citep{wilking90}, and allowed the extinction ${A_{\rm v}}$ to vary between 0 and 40 magnitudes. 

Before fitting these models to the data, the observed flux errors were reset to 10\% if they were below this value, to account for uncertainties not included in the photometric errors such as those due to variability. In addition to this, the 2MASS flux errors were reset to a factor of two, equivalent to down-weighting these data points by a factor of approximately 50 in the $\chi^2$ calculation, so that these data points would not over-dominate the fit. As shocked H$_2$ emission may be contributing to the measured flux in the 4.5\,$\mu$m IRAC band, we made this value an upper limit by rejecting models which had a 4.5\,$\mu$m flux higher than the measured flux. Each data point was also weighted by the distance in log wavelength between its two adjacent data points. This was done to allow the algorithm to search for a model which reproduced the shape of the entire SED, not only the sections with a high density of data. 

The models were also simultaneously fit to the six flux profiles shown in Figure \ref{fig:profiles}, measured from the four IRAC images and the 12.5 and 18.3\,$\mu$m images in \citet{de-buizer05}. The model flux profiles were found in the same way as the observed profiles, by summing the flux in the model images within a 15.1" wide strip aligned with and centred on the outflow axis, and normalising them to the total flux. When fit, the model profiles were allowed to be shifted by $\pm$2" (with the same shift for all profiles) around the centre of the observed profile to produce the best fit. This was necessary since the position of the central source could not be accurately determined from the images.

The combined $\chi^2$ for each model given as input to the genetic algorithm was calculated as

\begin{equation}
\chi^2= \chi_{\rm SED}^2 + \chi_{\rm profiles}^2
\end {equation} 

where $\chi_{\rm SED}^2$ is the best-fitting reduced $\chi^2$ value for the SED alone, and $\chi_{\rm profiles}^2$ is the overall reduced $\chi^2$ for the profile fits.

\subsection{The genetic algorithm}
\label{geneticshort}
In the context of SED modelling, genetic search algorithms \citep{holland75, goldberg89} have previously been used by \citet{hetem07} to model the protoplanetary discs of T Tauri and Herbig Ae/Be stars. Here we describe our algorithm, which was used in conjunction with the radiation transfer dust code described in Section \ref{dustcode} to search for the best fitting model for IRAS\,20126+4104.

Genetic algorithms are an alternative to traditional optimization methods, such as Markov chain Monte Carlo or simulated annealing. They move towards the optimum solution in an evolutionary manner, by generating generations of solutions or models, based on a given fitness statistic such as a $\chi^2$ value. Along with simulated annealing, they have the advantage that they do not become stuck in local minima, as their solution probabilistically tends towards the global optimum. In addition, genetic algorithms suit our specific problem because they involve producing or running several models at the same time. As each of the models was run with one million Monte Carlo photons \citep[actually energy packets,][]{bjorkman01}, taking approximately 30 minutes and 9 minutes per model to complete for the models with and without a disc respectively, this allows us to explore parameter space much more quickly than optimization codes which run one model at a time.

To create the first generation of models, $N$ models were generated by randomly sampling the parameters linearly or logarithmically within the ranges given in Table \ref{ranges}. The type of sampling (linear or log.) is also noted in the table. Here we took $N=100$.

The second generation consisted of $M$ models, where $M$ was some fraction of $N$. Here we chose $M$=$20$. Models in this and subsequent generations were computed by either the \textit{mutation} of one existing model, or the \textit{crossover} of two. We used \textit{tournament selection} to select which models were mutated or crossed. To create a tournament, a fraction of the models, $k$ (we chose $k=10\%$), were selected at random from the \textit{gene pool} of N models. The winner of each tournament was determined by first sampling a random number $q$ between 0 and 1. The probability $p$, between 0.5 and 1, which we chose to be $p=0.9$, was used to determine how likely a model is to win the tournament: if $q < p$, the model with the lowest $\chi^2$ value was selected. If not, and $q < p(p-1)$, then the second best model was selected, and so on such that if $q < p(p-1)^{n-1}$, where n is the nth model in the list ordered by $\chi^2$, the nth model wins the tournament. With this probabilistic selection, models that increase $\chi^2$ are occasionally promoted so that the evolution of the final model is not always toward a minimum. Therefore the algorithm allows escape from local minima. For mutations, one tournament was required; for crossovers, the pair selected for crossing resulted from two tournaments.

A mutation was carried out by selecting one of the varied parameters at random and resampling it from the original ranges given in Table~\ref{ranges}. A crossover was carried out by `crossing' the parameters of the two selected models, by randomly sampling  (either linearly or logarithmically) a value between the original parameters of the two models. In all generations following the first, 75\% of the models were the result of mutations, and 25\% were the result of crossovers. 

When the $M$ models of the next generation were added to the gene pool, the $M$ worst (highest $\chi^2$) models were removed so that the gene pool always contained $N$ models. All following generations were created using the same method as the second. The code was stopped when the best fitting model produced a less than 5\% reduction in $\chi^2$ over 20 generations.

\begin{table*}
 \centering
 \begin{minipage}[t]{5truein}
  \caption{Assumed ranges for model parameters as input to the genetic search algorithm.}
  \begin{tabular}{@{}llll@{}}
  \hline
 Parameter     & Description & Value/Range & Sampling \\
 \hline 
$M_{\star}$ & Stellar mass (M$_\odot$) & 5 -- 25 & logarithmic \\
$R^{\rm max}_{\rm env}$ & Envelope outer radius (au) & 15,000 -- 120,000 & logarithmic \\
$\theta_{\rm cav}$ & Cavity half-opening angle at 10$^{4}$\,au (degrees) & 5 -- 40 & linear \\
$i$ & Inclination w. r. t. the line of sight (degrees) & 75 -- 90 & linear \\
$M_{\rm disc}$ & Disc mass (M$_\odot$) & 0.1 -- 15 & logarithmic \\
$R_{\rm disc}$ & Disc or centrifugal radius (au) & 100 -- 10,000 & logarithmic \\
$\dot{M}_{\rm disc}$ & Disc accretion rate (M${}_\odot$ yr$^{-1}$) & $10^{-7}$ -- $10^{-2}$ & logarithmic \\
 $\dot{M}_{\rm env}$  & Envelope accretion rate ($M{}_\odot$ yr$^{-1}$) & $10^{-7}$ -- $10^{-2}$ & logarithmic \\
 $\rho_{\rm cav}$& Cavity ambient density (g\,cm${}^{-3}$)  & $10^{-21}$ -- $10^{-16}$ & logarithmic \\
 $R_{\star}$ & Stellar radius & Determined from stellar mass \\
 $T_{\star}$ & Stellar temperature &  Determined from stellar mass \\
\hline
\end{tabular}
\label{ranges}
\end{minipage}
\end{table*}

\section{Results}
\label{results}

The genetic code was run twice - firstly with the model parameter ranges given in Table~\ref{ranges} as input for all parameters, which we refer to as the `envelope plus disc' model, and secondly with the parameter ranges given in Table~\ref{ranges} for all parameters except the disc mass and disc accretion rate, $M_{\rm disc}$ and $\dot{M}_{\rm{disc}}$, which were set to zero and were therefore not treated as model parameters; we refer to this second run as the `envelope without disc' model. The envelope without disc model was run to ascertain if the SED and images could be adequately produced without a disc, i.e. a simpler model which had two fewer parameters. For the envelope without disc model, $R_{\rm disc}$ was instead referred to as the centrifugal radius of the envelope, $R_{\rm c}$, as these two parameters are interchangeable in the models.

The genetic search algorithm codes were stopped after 61 and 48 generations for the envelope plus disc and without disc models respectively, when the convergence criteria were reached. This corresponds to 1300 models or 631 CPU hours for the envelope plus disc model, and 1040 models or 152 CPU hours for the envelope without disc model. The resulting SEDs and profiles of these models are shown against the data in Figures~\ref{seddata} and \ref{fig:profiles}, and the model images are compared to the observed images in Figures \ref{IRACrgb} and \ref{fig:combined}. The parameters of the best-fitting models are also given in Table~\ref{bestfit}, along with the $\chi^2$ values for the SED-only fit, the fit to the image profiles, and the total $\chi^2$ for both the SED and profiles combined. We have included uncertainties on the parameters in Table~\ref{bestfit}, however these are not formal 1-$\sigma$ errors given by a $\Delta \chi^2$ of 1, as these confidence intervals would be unrealistically small for each of the fitted parameters, and would in most cases only contain the best fitting model. In addition, as the model is only a rough representation of reality, this uncertainty would probably underestimate the true error. Therefore, we have instead quoted in Table \ref{bestfit} the ranges corresponding to models with a \textit{reduced} $\chi^2$ less than $\chi^2_{\rm best}+20$, and have verified that models in this range still provided a reasonable fit to the data. 

\subsection{High signal-to-noise runs}
\begin{figure*}
\includegraphics[width=6.5in]{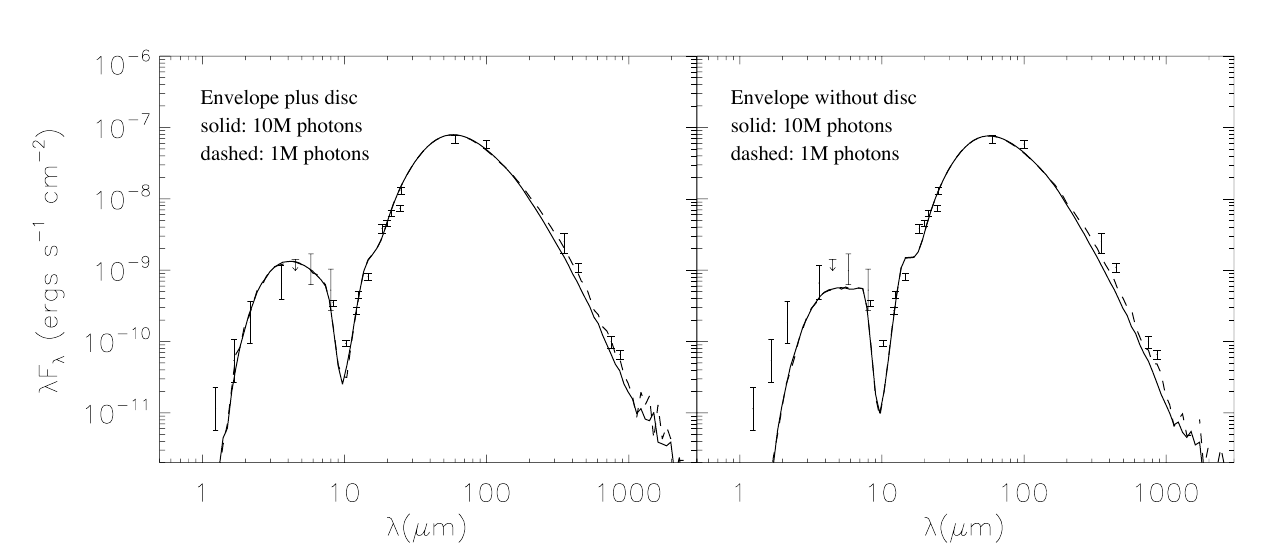}
 \caption{Left and right: the SEDs produced by a 10 million photon run and a one million photon run for the envelope plus and without disc models respectively.}
 \label{SED10m}
\end{figure*}

The model SEDs shown in Figure~\ref{seddata} are the result of running one million photon models. To obtain higher signal to noise in the SED and images, we ran the dust code with 10 million photons with the final set of parameters produced by the genetic code for both envelope plus disc and without disc models. The SEDs produced by the 10 million photon runs for both models are shown in the left and right panels of Figure \ref{SED10m} as solid lines, compared to the SEDs from the genetic code one million photon runs, shown as dashed lines. The SEDs agree well below $\sim$100$\mu$m, but are slightly offset at sub-mm wavelengths, due to the temperatures calculated by the dust code for the outer cells not being fully converged after one million photons, as few Monte Carlo photons have interacted in these cells. However, as the difference in SEDs is smaller than the uncertainties in the observed fluxes at the longest wavelengths, this is unlikely to have a significant impact on the results of the genetic algorithm. 

\subsection{Parameter $\chi^2$ surfaces}
\label{chi2surfacesection}

\begin{figure*}
\includegraphics[width=7.in]{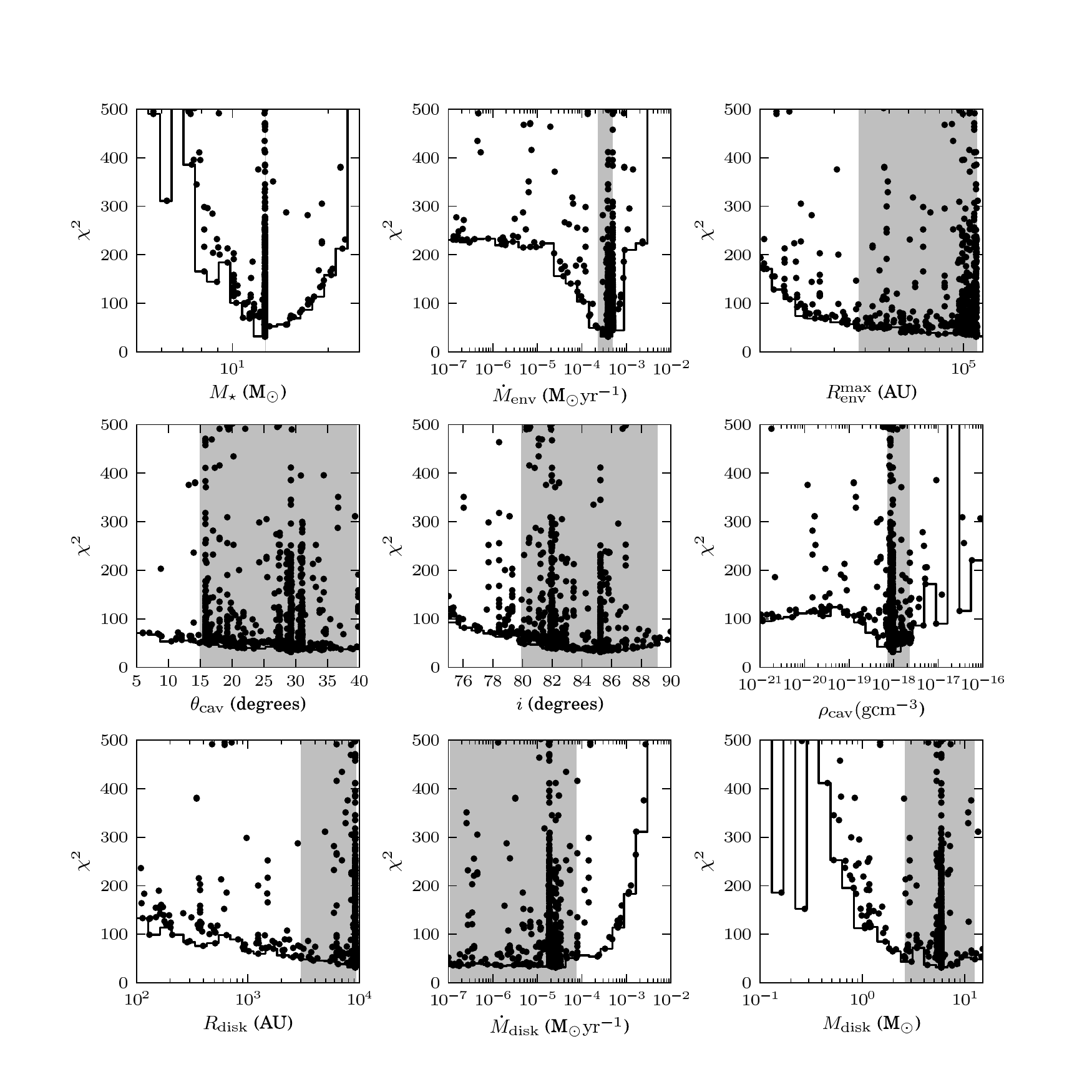}
 \caption{Plots of the $\chi^2$-value for models with reduced $\chi^2 < 500$ against the nine varied model parameters for the envelope plus disc model. The solid line shows a histogram of the minimum $\chi^2$-value in each bin, which provides a rough outline of the minimum $\chi^2$ surface. The grey shaded areas show the ranges of parameter values providing a ``good" fit (with reduced $\chi^2 - \chi_{\rm best}^2 <$20) given in Table~\ref{bestfit}.}
 \label{chi2surface}
\end{figure*}

\begin{figure*}
\includegraphics[width=7.in]{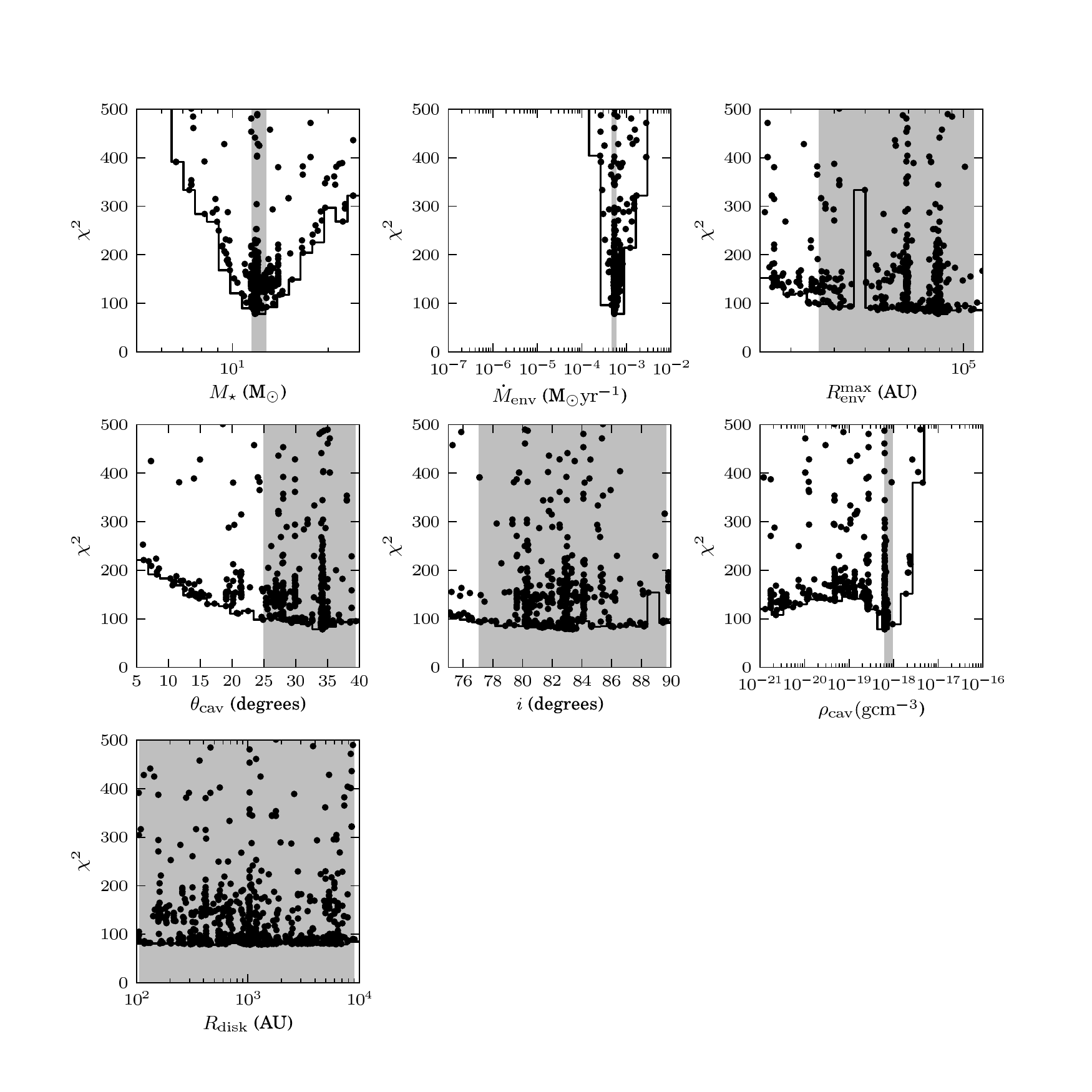}
 \caption{Plots of the $\chi^2$-value for models with reduced $\chi^2 < 500$ against the seven varied model parameters for the envelope without disc model. The solid line shows a histogram of the minimum $\chi^2$-value in each bin, which provides a rough outline of the minimum $\chi^2$ surface. The grey shaded areas show the ranges of parameter values providing a ``good" fit (with reduced $\chi^2 - \chi_{\rm best}^2 <$20) given in Table~\ref{bestfit}.}
 \label{chi2surface_nd}
\end{figure*}

To understand how well the parameters were determined,  we plotted for $\chi^2<500$ the reduced $\chi^2$-values of all the models run against each parameter, shown in Figures \ref{chi2surface} and \ref{chi2surface_nd}. As the first generation of the genetic code was sampled uniformly throughout the chosen ranges, we were therefore able to adequately sample parameter space and recover an approximation of the minimum $\chi^2$ surface. Therefore, the solid lines in Figures \ref{chi2surface} and \ref{chi2surface_nd} also show histograms of the minimum $\chi^2$ model in each bin. In addition, the grey shaded areas show the ranges of parameter values providing a ``good" fit (with reduced $\chi^2 - \chi_{\rm best}^2 <$20) given in Table \ref{bestfit}. From these Figures, we can understand to what extent the observed SED and profiles of IRAS\,20126+4104 constrain the properties of the source. 

For the envelope plus disc model, Figure \ref{chi2surface} shows that the stellar mass $M_{\star}$, the envelope accretion rate $\dot{M}_{\rm env}$, and the cavity ambient density $\rho_{\rm cav}$ are well-determined within the chosen ranges, having sharp minimum $\chi^2$ surfaces. The other parameters are not as tightly constrained, however the envelope radius $R^{\rm max}_{\rm env}$, the disc radius $R_{\rm disc}$ and the disc mass $M_{\rm disc}$ tend to have better fits at higher values. Also, the disc accretion rate $\dot{M}_{\rm disc}$ is not well fit by models with values above $\sim10^{-4}$ M$_{\odot}$yr$^{-1}$. The least well-determined parameter is the cavity half-opening angle $\theta_{cav}$.

For the envelope without disc model, Figure \ref{chi2surface_nd} shows that the parameters are constrained similarly to those in the envelope plus disc model. However, differences include that envelope accretion rates below $\sim$10$^{-4}$\,M$_{\odot}$yr$^{-1}$ are strongly ruled out, there is a deeper minimum towards higher values of the cavity half-opening angle, and no models exist with $\chi^{2}<500$ for values of the cavity ambient density greater than 5$\times$10$^{-18}$\,g\,cm$^{-3}$. In addition, both the inclination and centrifugal radius of the envelope $R_{\rm c}$ are very poorly constrained for the envelope without disc model. 

\begin{figure*}
\includegraphics[width=7.in]{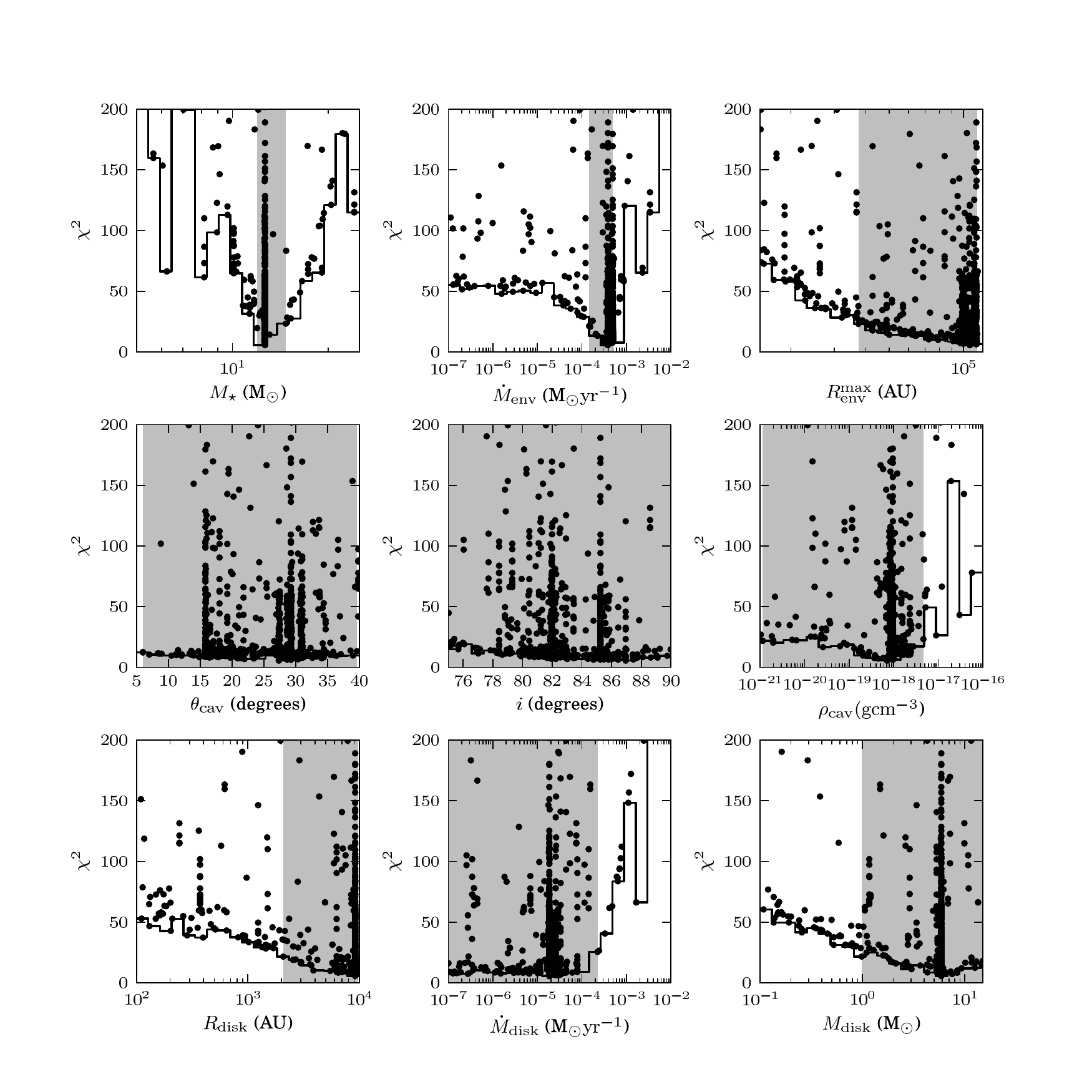}
 \caption{Plots of the SED-only $\chi_{_{\rm SED}}^2$-value for models with reduced $\chi_{_{\rm SED}}^2 < 200$ against the nine varied model parameters for the envelope plus disc model. The solid line shows a histogram of the minimum $\chi_{_{\rm SED}}^2$-value in each bin, which provides a rough outline of the minimum $\chi_{_{\rm SED}}^2$ surface. The grey shaded areas show the ranges of parameter values providing a ``good" fit (with reduced $\chi_{_{\rm SED}}^2 - \chi_{_{\rm SED, best}}^2 <$20).}
 \label{chi2surface_wd_sed}
\end{figure*}

\begin{figure*}
\includegraphics[width=7.in]{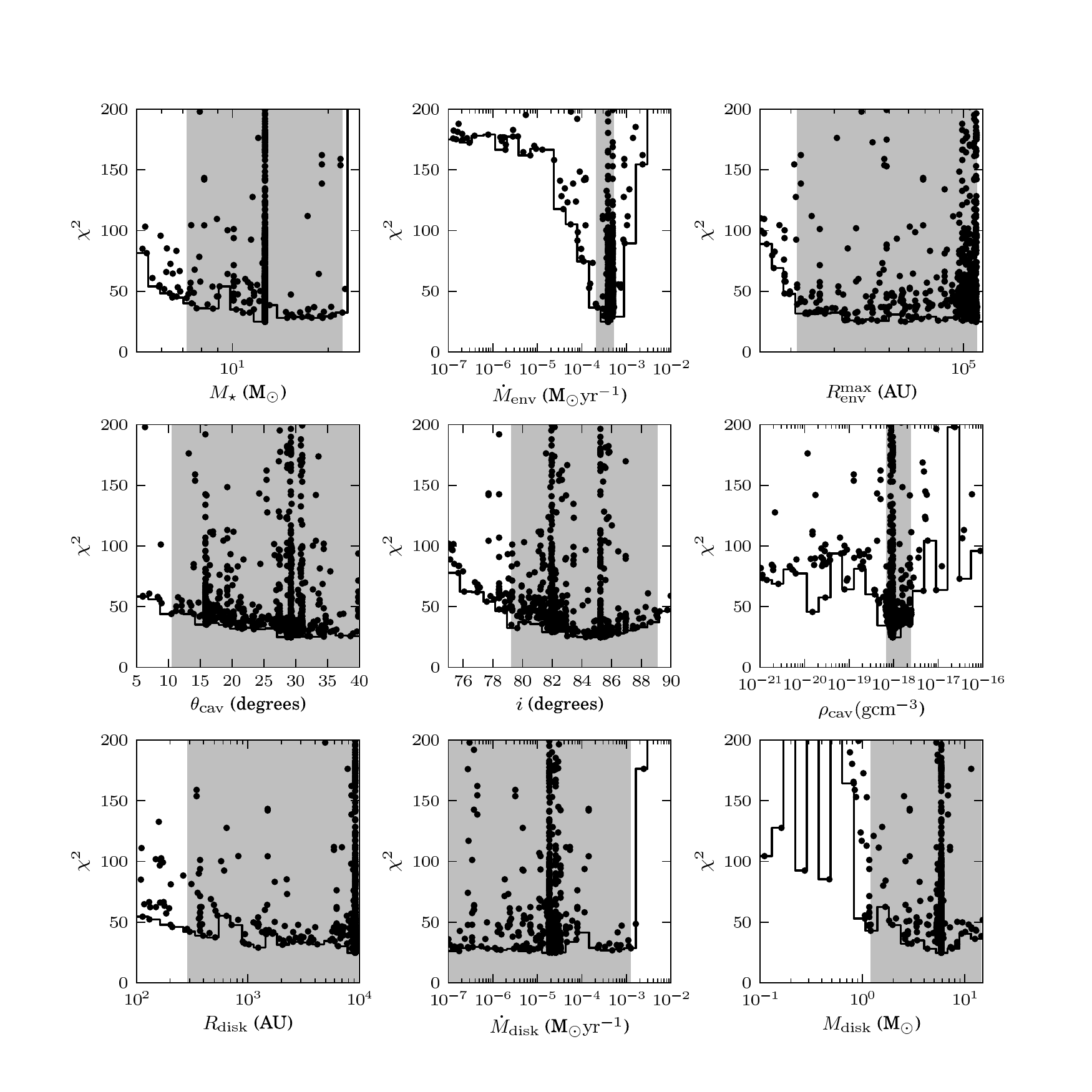}
 \caption{Plots of the profiles-only $\chi_{\rm profiles}^2$-value for models with reduced $\chi_{\rm profiles}^2 < 200$ against the nine varied model parameters for the envelope plus disc model. The solid line shows a histogram of the minimum $\chi_{\rm profiles}^2$-value in each bin, which provides a rough outline of the minimum $\chi_{\rm profiles}^2$ surface. The grey shaded areas show the ranges of parameter values providing a ``good" fit (with reduced $\chi_{\rm profiles}^2 - \chi_{\rm profiles,best}^2 <$20).}
 \label{chi2surface_wd_prof}
\end{figure*}

For the \textit{envelope plus disc model}, we also include plots of the reduced $\chi^2$ against each parameter for all the models with $\chi^{2}<200$ for the SED-only fit (Figure~\ref{chi2surface_wd_sed}) and for the profiles-only fit (Figure~\ref{chi2surface_wd_prof}). This allows us to see more clearly whether the SED or profiles are constraining a given parameter. For instance, it can be seen that the stellar mass, envelope outer radius and disc outer radius are better constrained by the SED, and the cavity opening angle, cavity density as well as the viewing angle $i$ are better constrained by the profiles. The remaining parameters are similarly constrained by both the SED and profiles.

\begin{table*}
 \centering
 \begin{minipage}[t]{5.5truein}
  \caption{Parameters of the genetic algorithm best-fitting models}
  \begin{tabular}{@{}llll@{}}
  \hline
Parameter     & Description & Value for envelope & Value for envelope \\
& & plus disc model& without disc model \\
 \hline 
$M_{\star}$ & Stellar mass (M$_\odot$) & 12.7$\pm0.0$ & 11.8$\pm_{0.3}^{1.0}$\\ [+0.02in]
$R^{\rm max}_{\rm env}$ & Envelope outer radius (au) & 113,000$\pm_{75000}^{1000}$ & 79,100$\pm_{53200}^{31100}$\\ [+0.02in]
$\theta_{\rm cav}$ & Cavity half-opening angle at 10$^{4}$\,au (degrees) & 29.3$\pm_{14.4}^{10.4}$  & 34.2$\pm_{9.3}^{5.3}$ \\ [+0.02in]
$i$ & Inclination w. r. t. the line of sight (degrees) & 85.2$\pm_{5.3}^{3.9}$ & 83.4$\pm6.3$ \\ [+0.02in]
$M_{\rm disc}$ & Disc mass (M$_\odot$) & 5.90$\pm_{3.29}^{6.68}$ & ... \\ [+0.02in]
$R_{\rm disc}$ or $R_{\rm c}$ & Disc or centrifugal radius (au) & 9200$\pm_{6200}^{300}$ & 1040$\pm_{940}^{8040}$ \\ [+0.02in]
$\dot{M}_{\rm disc}$ & Disc accretion rate (M${}_\odot$ yr$^{-1}$) & 2.58$\pm_{2.57}^{4.90}$$\times 10^{-5}$ & ... \\ [+0.02in]
 $\dot{M}_{\rm env}$  & Envelope accretion rate (M${}_\odot$ yr$^{-1}$) & 3.86$\pm_{1.61}^{1.06}$$\times 10^{-4}$  & 5.35$\pm_{0.73}^{0.66}$$\times 10^{-4}$ \\ [+0.02in]
 $\rho_{\rm cav}$& Cavity ambient density (g\,cm${}^{-3}$)  & 9.55$\pm_{2.48}^{13.48}$$\times 10^{-19}$ & 6.27$\pm_{0.01}^{3.22}$$\times 10^{-19}$ \\ [+0.02in]
 $R_{\star}$ & Stellar radius (R$_{\odot}$) & 4.53 & 4.34 \\ [+0.02in]
 $T_{\star}$ & Stellar temperature (K) & 28,700 & 27,800 \\ [+0.02in]
 $\chi^2$(SED) & $\chi^2$ value for the SED fit only & 7.0 & 24.6 \\ [+0.02in]
 $\chi^2$(profiles) & $\chi^2$ value for the fit to the profiles only & 24.9 & 53.8 \\ [+0.02in]
 $\chi^2$(Total) & Total $\chi^2$ for both SED and profiles & 31.9 & 78.4 \\ [+0.02in]
 \hline
\end{tabular}
Note: the uncertainties on the parameter values are given by the range of models with a reduced $\chi^2 - \chi_{\rm best}^2 <20$.
\label{bestfit}
\end{minipage}
\end{table*}

\subsection{Comparison to the observed SED}

We find that the minimum reduced $\chi^2$ for the best-fitting envelope plus disc model is 31.9, and the reduced $\chi^2$ for the best-fitting envelope without disc model is 78.4, from which we infer that the model with a disc provides a better fit to the SED and profiles. For the SED, this can be seen in Figure \ref{seddata}, where the envelope without disc model does not reproduce the observed near- and mid-IR fluxes. 

\subsection{Comparison to the observed profiles and images}
\label{modelimages}

Here we compare the model profiles and images to those observed. Figure \ref{fig:profiles} presents the model profiles for the envelope plus disc and without disc models. The left and right panels of Figure \ref{IRACrgb} also present three-colour RGB model images of the IRAC emission toward IRAS\,20126+4104: red: 8\,$\mu$m, green: 5.8\,$\mu$m, and blue: 3.6\,$\mu$m, for both envelope plus disc and without disc models. 

Both the IRAC images in Figure \ref{IRACrgb} and the profiles in Figure \ref{fig:profiles} show that while both models are able to reproduce the observed angular size of the mid-IR emission, the envelope without disc model is not able to fully reproduce the shape of the IRAC emission, as its 8\,$\mu$m model image is too centrally peaked.

Figure \ref{fig:combined} also presents the observed and corresponding model images at the wavelengths and spatial resolutions observed in \citet{sridharan05}, \citet{de-buizer07} and \citet{de-wit09}: K band, 12.5$\mu$m, 18.3$\mu$m and 24.5$\mu$m. 
All model images in Figures \ref{IRACrgb} and \ref{fig:combined} were roughly aligned to the observed images using the centre of the dark lane in the observed and envelope without disc model 12.5$\mu$m images, and were rotated clockwise by 56.5$^{\circ}$ to reproduce the observed position angle. 

The morphology of the K band model images (top left and right panels of Figure \ref{fig:combined}) are able to roughly reproduce the two outflow cavities imaged by \citet{sridharan05}, shown in the top middle panel of Figure~\ref{fig:combined}. Comparing the mid-IR model images shown at the remaining three wavelengths, we see that the mid-IR model images are strongly affected by the presence of a disc, specifically in the separation of the two outflow cavity emission lobes. In Figure \ref{fig:combined} we also note that the mid-IR images produced by the envelope without disc model reproduce the observed mid-IR images much better than those of the model with a disc, and note that the lack of a dark lane in the observed 24.5$\mu$m image is also reproduced by the envelope without disc model, while a dark lane is still seen in the shorter wavelength images. Here the shadowing produced by a rotationally flattened envelope appears to be adequate to match the observed dark lane at 12.5 and 18.3$\mu$m.

However, there is further emission to the north west in the observed 12.5, 18.3 and 24.5\,$\mu$m images, cut off by another `dark lane', which cannot be reproduced by our models. Several possibilities which may explain this image morphology include: 

\begin{enumerate}
\item There is a disc toward IRAS\,20126+4104, in agreement with the observed SED and IRAC images. However, the disc is truncated or removed inside the radius delineated by the dark lane between the central emission and the northwest emission (at a radius of $\sim$1.1" or $\sim$1800\,au), so that the emission within this radius resembles the model without a disc. Mechanisms which would truncate the disc include stellar winds, ionization and heating, which would therefore imply a difference between IRAS\,20126+4104 and its lower-mass counterparts. 

\item Precession of the outflow axis with time, so that the outflow cavity at smaller radii is not aligned with the outflow cavity at larger radii, explaining the observed discontinuity in the images at a radius of $\sim$1800\,au. In fact, there is evidence that the outflow IRAS\,20126+4104 is precessing \citep{shepherd00, cesaroni05b}.

\item The emission within $\sim$1800\,au is produced by a different young star to that producing the larger scale infrared emission. A companion is expected, as the binary fraction of massive stars is higher than their low-mass counterparts. In addition, a binary companion may explain the outflow precession. 

\end{enumerate}

\subsection{The effect of different dust properties}
\begin{figure*}
\includegraphics[width=4.in]{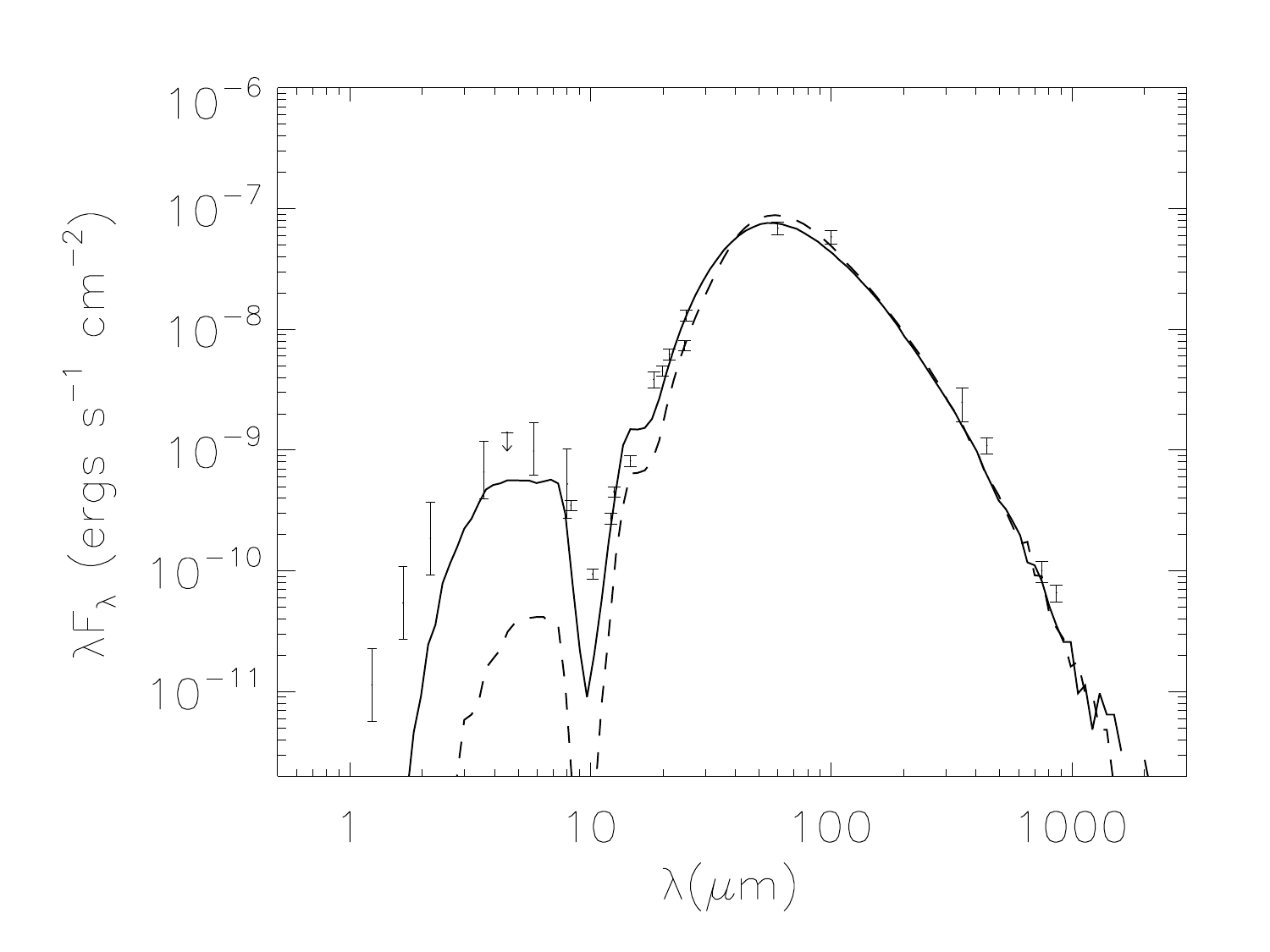}
 \caption{The SED of the best-fitting envelope without disc model using the dust properties given by KMH (solid line) and Draine (dashed line).}
 \label{SEDalbedo}
\end{figure*}

To show the effect different dust properties would have on the results of the genetic code, and whether different dust properties could have a similar effect to the presence of a disc, Figure~\ref{SEDalbedo} compares the SEDs of the envelope without disc model for two different sets of realistic dust properties. The solid line represents the SED produced using the same dust properties as those used for the genetic code, taken from KMH, and the dashed line shows the resultant SED using dust properties from \citeauthor{draine03a} (\citeyear{draine03a,draine03b}, hereafter Draine). The Draine dust model reproduces the Milky-Way extinction curve for an $R_v$=5.5 (case A with C/H = 30 ppm) using the grain size distribution from \citet{weingartner01}. The composition is a mixture of carbonaceous grains and amorphous silicate grains, and this model has fewer small grains than models that fit a typical interstellar extinction curve (such as the KMH model we use in this paper). This model therefore has a higher albedo at near-IR ($<3\,\mu$m) wavelengths. Figure~\ref{SEDalbedo} shows that the Draine SED falls short of the KMH SED in the near- and mid-IR regime, and hence a different set of realistic dust properties with a higher albedo at near-IR wavelengths is not be able to reproduce the effect of the addition of a disc to the geometry (such as shown in Figure~\ref{seddata}). This is due to the fact that although the increased average grain size increases the albedo in the near-IR regime, it also simultaneously increases the extinction, which therefore reduces the emergent flux.

\subsection{Comparison with molecular line modelling results of KZ10}

In this Section, we compare the properties of IRAS\,20126+4104 found using two different approaches. The first has modelled the profiles of molecular lines observed toward IRAS\,20126+4104 in the millimetre (KZ10), and the second has modelled the near-IR to sub-millimetre continuum SED and images (this work). As both studies find that their models require a disc to adequately fit the data, Table \ref{tablecompare} compares the envelope plus disc model from this work and the parameters of the best-fitting disc model of KZ10. 

In Table \ref{tablecompare}, we see that both models find a similar stellar mass, disc mass and disc radius, as well as disc accretion rate, which was assumed in the KZ10 model to be equal to the envelope accretion rate. 

The envelope outer radius was not fit in the KZ10 model, as their model only extended out to a radius of $\sim$26,400\,au or 0.128\,pc. In addition, the inclination was assumed to be 90 degrees, similar to our fitted value of 85.2 degrees. As in our model, the stellar radius and temperature in the KZ10 model were determined from the stellar mass. 

The envelope accretion rate, $\dot{M}_{\rm env}$, the envelope density at the disc radius, $\rho_0$, and the total mass within a radius of 26,400\,au, $M_{\rm tot}$(0.128\,pc), all have $\sim$3-5 times larger values for our envelope plus disc model than the KZ10 disc model. This discrepancy may be explained by the fact that our model probes the envelope out to larger radii, and therefore can more accurately determine the density in the envelope. Alternatively, since the mass density in KZ10 depends on an assumed NH$_3$ abundance, this discrepancy may also be resolved by increasing the assumed NH$_3$ abundance by a factor of $\sim$3-5.

We fit the temperature in the midplane calculated by the dust code for our best-fitting envelope plus disc model with a power-law at radii greater than 10,000\,au, and found that the temperature varied here as $T \propto R^{-0.34}$. In comparison, the temperature in the KZ10 disc model was found to decrease more steeply than $T \propto R^{-2/3}$, corresponding to $p<-1$, where $p$ was used to define the temperature power-law exponent in the envelope as $T \propto R^{-2/(4+p)}$. 

The disc density ratio, $A_{\rho}$ is the ratio of the disc density to the envelope density at the disc radius (equation 9 of KZ10). Although the two values of $A_{\rho}$ given in Table \ref{tablecompare} differ noticeably, this is due to different values of the envelope density $\rho_0$, while the disc densities at $R_{\rm disc}$ are similar. The disc temperature factor, $B_T$, is the coefficient in the expression describing the disc temperature given in equation 12 of KZ10. This parameter does not have an equivalent in our model, as the temperatures for the envelope plus disc geometry were solved for by the dust code.

Finally, comparing the dynamics of the two models, we see that the angular momentum and the velocity at the disc radius have similar values for both models.

\begin{table*}
 \centering
 \begin{minipage}[t]{6.5truein}
  \caption{Comparison of parameters of genetic best-fitting envelope plus disc model with those from KZ10}
  \begin{tabular}{@{}llll@{}}
  \hline
  Parameter     & Description & Value for envelope plus & Value for model with disc \\
& & disc model (this work) & (KZ10) \\
 \hline 
$M_{\star}$ & Stellar mass (M$_\odot$) & 12.7$\pm0.0$ & 10.7 \\ [+0.02in]
$R^{\rm max}_{\rm env}$ & Envelope outer radius (au) & 113,000$\pm_{75000}^{1000}$ & $>$26,400 (assumed) \\ [+0.02in]
$\theta_{\rm cav}$ & Cavity half-opening angle at 10$^{4}$\,au (degrees) & 29.3$\pm_{14.4}^{10.4}$ & ...\\ [+0.02in]
$i$ & Inclination w. r. t. the line of sight (degrees) & 85.2$\pm_{5.3}^{3.9}$ & 90 (assumed) \\ [+0.02in]
$M_{\rm disc}$ & Disc mass (M$_\odot$) & 5.90$\pm_{3.29}^{6.68}$ & 2.5 \\ [+0.02in]
$R_{\rm disc}$ & Disc or centrifugal radius (au) & 9200$\pm_{6200}^{300}$ & 6900 \\ [+0.02in]
$\dot{M}_{\rm disc}$ & Disc accretion rate (M${}_\odot$ yr$^{-1}$) & 2.58$\pm_{2.57}^{4.90}$$\times 10^{-5}$ & ($7.6\times 10^{-5}$)  \\ [+0.02in]
 $\dot{M}_{\rm env}$  & Envelope accretion rate ($M{}_\odot$ yr$^{-1}$) & 3.86$\pm_{1.61}^{1.06}$$\times 10^{-4}$ & $7.6\times 10^{-5}$ \\ [+0.02in]
 $\rho_{\rm cav}$& Cavity ambient density (g\,cm${}^{-3}$)  & 9.55$\pm_{2.48}^{13.48}$$\times 10^{-19}$ & ...\\ [+0.02in]
 $R_{\star}$ & Stellar radius ($R_\odot$) & 4.53 & 4.8 \\ [+0.02in]
 $T_{\star}$ & Stellar temperature ($K$) & 28,700 & 19,000 \\ [+0.02in]
 $\rho_0$ & Env. density at $R_d$ (cm$^{-3}$) & $2.39\times 10^5$  & $7.9\times 10^4$  	\\ [+0.02in]
 $p$ & Temperature power law exponent ($T \propto R^{-2/(4+p)}$) & 1.8 ($>$10,000\,au, $T \propto R^{-0.34}$) & $< -1$ ($T \propto R^{-2/3}$) \\ [+0.02in]
 $\Gamma$ & Angular momentum	(au kms$^{-1}$) & 10,200 & 8100 	\\[+0.02in]
 $A_{\rho}$ & Disc density ratio  & 0.792 & 5.1 \\ [+0.02in]
 $B_{T}$ & Disc temperature factor & ... & 15.0 \\ [+0.02in]
 $v_k$ & Velocity at $R_d$ (kms$^{-1}$)	& 1.11 & 1.2			\\ [+0.02in]
 $M_{\rm tot}$(0.128\,pc) & Total mass within 0.128\,pc (M$_{\odot}$) & 49.7 & 12.6 \\ [+0.02in]
 $M_{\rm tot}(R^{\rm max}_{\rm env})$ & Total mass within $R^{\rm max}_{\rm env}$ (M$_{\odot}$) & 649 & ... \\ [+0.02in]
 \hline
\end{tabular}
\label{tablecompare}
\end{minipage}
\end{table*}

\subsection{Properties of the disc}

Here we discuss the properties of the disc around IRAS\,20126+4104 suggested by our results. The disc radius $R_{\rm{disc}}$ for the envelope plus disc model was found by the genetic code to be 9200$\pm_{6200}^{300}$\,au. For a Keplerian velocity of 1.11\,kms$^{-1}$ at $R_{\rm{disc}}$, the outer radius of the disc would therefore take $\sim$3$\times10^{5}$yr to complete one orbit. As this is on the order of the time massive stars spend in their accretion phase, it appears unlikely that the outer regions of the disc will have had the time to reach centrifugal or hydrostatic equilibrium, which occur on approximately the same time-scales. However, our results show that, in addition to the rotationally flattened envelope, extra mass arranged in some form of flattened structure is required, for radii up to 9200\,au, to reproduce the observed SED and profiles. 

\begin{figure}
\includegraphics[width=3.5in]{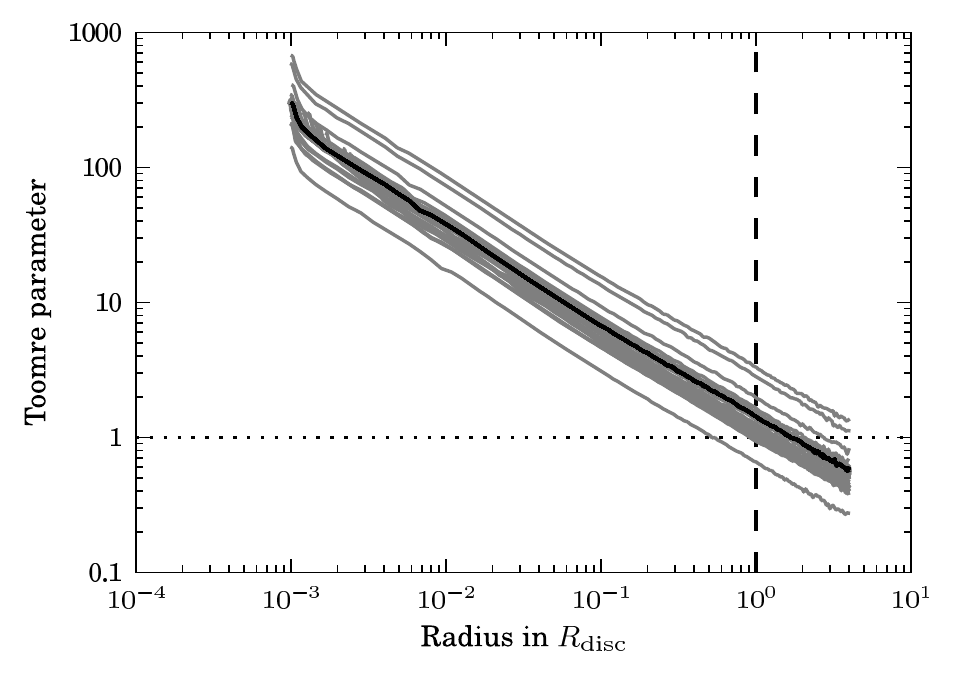}
 \caption{The Toomre parameter in the disc midplane as a function of disc radius (in $R_{\rm disc}$) for the best-fitting envelope with disc model, shown as a black line, and for the models with reduced $\chi^2 - \chi_{\rm best}^2 <$20, shown as grey lines.}
 \label{toomre}
\end{figure}

As the dust code calculates the temperature of the envelope and disc geometry, we were able to calculate the Toomre Q parameter \citep{toomre64} - a measure of stability towards fragmentation - in the midplane of the disc as a function of radius,
\begin{equation}
Q=\frac{c_s \Omega}{\pi G \Sigma}
\end{equation}
where $c_s$ is the speed of sound in the gas, $\Omega$ is the angular velocity of the disc, $G$ is the gravitational constant and $\Sigma$ is the surface density of the disc. We found the surface density as a function of radius by integrating equation \ref{discdensity} over the disc height $z$. To find the sound speed, we assumed an ideal gas composed of molecular hydrogen, 
\begin{equation}
c_s=\sqrt{\frac{kT}{2.3 m_H}}
\end{equation}
where $k$ is the Boltzmann constant, $T$ is the temperature calculated by the dust code and $2.3 m_H$ is the average molecular mass of the gas. The Toomre parameter in the disc midplane as a function of disc radius for the best-fitting envelope with disc model is shown as a black line in Figure \ref{toomre}. From this, we see that the disc is stable ($Q>1$), and only reaches a value less than unity at radii greater than the disc radius. Thus the high temperatures and low densities within the disc stabilise it against local fragmentation. We also show as grey lines the Toomre parameter for the models with reduced $\chi^2 - \chi^2_{\rm best} < 20$, which, apart from one model, lie close to or above 1 at the disc radius. Therefore we conclude that across the range of uncertainty in the disc properties, the disc around IRAS\,20126+4104 is still stable to fragmentation.

\section{Conclusions}
\label{conclusions}

In order to determine whether the standard model of low-mass star formation can also apply to massive stars, we have modelled the near-IR to sub-millimetre SED and several mid-IR images of the accreting embedded massive star IRAS\,20126+4104, using model geometries which are commonly used to describe forming low-mass stars.

We have used a Monte Carlo radiative transfer dust code to model the continuum absorption, emission and scattering through two scaled-up azimuthally symmetric dust geometries, the first consisting of a rotationally flattened envelope with outflow cavities, and the second which also includes a flared disc. 

To find the best-fitting set of model parameters, we used a genetic algorithm to search parameter space, within parameter ranges which bracketed previous results or were physically plausible. This also allowed us to produce minimum $\chi^2$ surfaces for each parameter, from which we could infer how well the SED and image profiles constrained them. We found that the two parameters most constrained by the data were the stellar mass and envelope accretion rate, which were determined to be $\sim$13\,$\rm{M}_{\odot}$ and $\sim4\times10^{-4}\rm{\,M}_{\odot}$yr$^{-1}$ respectively for the envelope plus disc model, and $\sim$12\,$\rm{M}_{\odot}$ and $\sim5\times10^{-4}\rm{\,M}_{\odot}$yr$^{-1}$ for the envelope without disc model. These two parameters are actually proxies for other more directly determined properties of the source - the stellar mass in the models is a proxy for the stellar luminosity, and the envelope accretion rate is a proxy for the total mass in the envelope, which is constrained by the millimetre fluxes. The stellar luminosities for both the envelope plus and without disc models respectively are $1.3 \times10^{4}$ and $1.0 \times10^{4}$\,L$_{\sun}$, and the total masses within $R^{\rm max}_{\rm env}$ are 649 and 375\,M$_{\sun}$. The best-fitting values of the remaining parameters are given in Table~\ref{bestfit}.

Our results show that the envelope plus disc model reproduces the observed SED and images better than the envelope without disc model, although the model without a disc appears to better-reproduce the morphology of the mid-IR emission within a radius of $\sim$1800\,au. We have outlined several possible causes of this discontinuity, such as inner disc truncation or outflow precession. It may be that the observed mid-IR images, which cannot be fully explained by the envelope plus disc model, are showing the effect of the radiation of a massive star on its accreting material. However, future observations and modelling are needed to determine whether precession and/or a young protostellar companion are more likely explanations.

Comparing our results to those of KZ10, we note that both studies find that a model with a disc reproduces the observations more successfully than one without. Although both studies modelled completely different observations using different techniques, we find that our best-fitting model parameters are similar to those found by KZ10. While we find a higher envelope accretion rate than KZ10, and therefore a higher envelope density and total envelope mass, this difference may be due to the fact that the masses determined from the molecular line modelling depend on the assumed molecular abundances. For example, increasing the assumed molecular abundances by a factor of 3-5 would bring these values into agreement. In addition, as the observed SED is more sensitive to emission from the entire envelope, not only the material within a radius of 26,400\,au modelled by KZ10, we were able to probe the mass in the envelope more accurately.

Our best fitting envelope plus disc model has a disc radius of 9200\,au. As at this radius the disc requires $\sim$3$\times10^{5}$\,yr to complete one orbit, we find it is unlikely that the outer regions of the disc have had time to reach centrifugal or hydrostatic equilibrium. However, by using the temperature along the disc midplane found by the dust code, we calculate that this disc would be stable to local fragmentation.

\section*{Acknowledgments}
The authors would like to thank T.K. Sridharan, J. De Buizer, and W. de Wit for providing electronic versions of their infrared images to include in this work, and also the anonymous referee, who provided many insightful comments which significantly improved the paper.

\bibliography{}

\begin{thebibliography}{}

\bibitem[\protect\citeauthoryear{{Adams}, {Lada} \& {Shu}}{{Adams}
  et~al.}{1987}]{adams87}
{Adams} F.~C.,  {Lada} C.~J.,    {Shu} F.~H.,  1987, \apj, 312, 788

\bibitem[\protect\citeauthoryear{{Akeson}, {Walker}, {Wood}, {Eisner}, {Scire},
  {Penprase}, {Ciardi}, {van Belle}, {Whitney} \& {Bjorkman}}{{Akeson}
  et~al.}{2005}]{akeson05}
{Akeson} R.~L.,  {Walker} C.~H.,  {Wood} K.,  {Eisner} J.~A.,  {Scire} E.,
  {Penprase} B.,  {Ciardi} D.~R.,  {van Belle} G.~T.,  {Whitney} B.,
  {Bjorkman} J.~E.,  2005, \apj, 622, 440

\bibitem[\protect\citeauthoryear{{Bjorkman} \& {Wood}}{{Bjorkman} \&
  {Wood}}{2001}]{bjorkman01}
{Bjorkman} J.~E.,  {Wood} K.,  2001, \apj, 554, 615

\bibitem[\protect\citeauthoryear{{Calvet} \& {Gullbring}}{{Calvet} \&
  {Gullbring}}{1998}]{calvet98}
{Calvet} N.,  {Gullbring} E.,  1998, \apj, 509, 802

\bibitem[\protect\citeauthoryear{{Castelli} \& {Kurucz}}{{Castelli} \&
  {Kurucz}}{2004}]{castelli04}
{Castelli} F.,  {Kurucz} R.~L.,  2004, ArXiv Astrophysics e-prints

\bibitem[\protect\citeauthoryear{{Cesaroni}, {Felli}, {Jenness}, {Neri},
  {Olmi}, {Robberto}, {Testi} \& {Walmsley}}{{Cesaroni}
  et~al.}{1999}]{cesaroni99b}
{Cesaroni} R.,  {Felli} M.,  {Jenness} T.,  {Neri} R.,  {Olmi} L.,  {Robberto}
  M.,  {Testi} L.,    {Walmsley} C.~M.,  1999, \aap, 345, 949

\bibitem[\protect\citeauthoryear{{Cesaroni}, {Felli}, {Testi}, {Walmsley} \&
  {Olmi}}{{Cesaroni} et~al.}{1997}]{cesaroni97}
{Cesaroni} R.,  {Felli} M.,  {Testi} L.,  {Walmsley} C.~M.,    {Olmi} L.,
  1997, \aap, 325, 725

\bibitem[\protect\citeauthoryear{{Cesaroni}, {Felli} \& {Walmsley}}{{Cesaroni}
  et~al.}{1999}]{cesaroni99}
{Cesaroni} R.,  {Felli} M.,    {Walmsley} C.~M.,  1999, \aaps, 136, 333

\bibitem[\protect\citeauthoryear{{Cesaroni}, {Neri}, {Olmi}, {Testi},
  {Walmsley} \& {Hofner}}{{Cesaroni} et~al.}{2005}]{cesaroni05b}
{Cesaroni} R.,  {Neri} R.,  {Olmi} L.,  {Testi} L.,  {Walmsley} C.~M.,
  {Hofner} P.,  2005, \aap, 434, 1039

\bibitem[\protect\citeauthoryear{{Chieffi} \& {Straniero}}{{Chieffi} \&
  {Straniero}}{1989}]{chieffi89}
{Chieffi} A.,  {Straniero} O.,  1989, \apjs, 71, 47

\bibitem[\protect\citeauthoryear{{D'Alessio}, {Canto}, {Calvet} \&
  {Lizano}}{{D'Alessio} et~al.}{1998}]{dalessio98}
{D'Alessio} P.,  {Canto} J.,  {Calvet} N.,    {Lizano} S.,  1998, \apj, 500,
  411

\bibitem[\protect\citeauthoryear{{De Buizer}}{{De Buizer}}{2007}]{de-buizer07}
{De Buizer} J.~M.,  2007, \apjl, 654, L147

\bibitem[\protect\citeauthoryear{{De Buizer}, {Radomski}, {Telesco} \&
  {Pi{\~n}a}}{{De Buizer} et~al.}{2005}]{de-buizer05}
{De Buizer} J.~M.,  {Radomski} J.~T.,  {Telesco} C.~M.,    {Pi{\~n}a} R.~K.,
  2005, \apjs, 156, 179

\bibitem[\protect\citeauthoryear{{de Wit}, {Hoare}, {Fujiyoshi}, {Oudmaijer},
  {Honda}, {Kataza}, {Miyata}, {Okamoto}, {Onaka}, {Sako} \& {Yamashita}}{{de
  Wit} et~al.}{2009}]{de-wit09}
{de Wit} W.~J.,  {Hoare} M.~G.,  {Fujiyoshi} T.,  {Oudmaijer} R.~D.,  {Honda}
  M.,  {Kataza} H.,  {Miyata} T.,  {Okamoto} Y.~K.,  {Onaka} T.,  {Sako} S.,
  {Yamashita} T.,  2009, \aap, 494, 157

\bibitem[\protect\citeauthoryear{{Draine}}{{Draine}}{2003a}]{draine03a}
{Draine} B.~T.,  2003a, \araa, 41, 241

\bibitem[\protect\citeauthoryear{{Draine}}{{Draine}}{2003b}]{draine03b}
{Draine} B.~T.,  2003b, \apj, 598, 1017

\bibitem[\protect\citeauthoryear{{Edris}, {Fuller}, {Cohen} \& {Etoka}}{{Edris}
  et~al.}{2005}]{edris05}
{Edris} K.~A.,  {Fuller} G.~A.,  {Cohen} R.~J.,    {Etoka} S.,  2005, \aap,
  434, 213

\bibitem[\protect\citeauthoryear{{Goldberg}}{{Goldberg}}{1989}]{goldberg89}
{Goldberg} D.~E.,  1989, {Genetic algorithms in search, optimization and
  machine learning}

\bibitem[\protect\citeauthoryear{{Hetem} \& {Gregorio-Hetem}}{{Hetem} \&
  {Gregorio-Hetem}}{2007}]{hetem07}
{Hetem} A.,  {Gregorio-Hetem} J.,  2007, \mnras, 382, 1707

\bibitem[\protect\citeauthoryear{{Hofner}, {Cesaroni}, {Olmi},
  {Rodr{\'{\i}}guez}, {Mart{\'{\i}}} \& {Araya}}{{Hofner}
  et~al.}{2007}]{hofner07}
{Hofner} P.,  {Cesaroni} R.,  {Olmi} L.,  {Rodr{\'{\i}}guez} L.~F.,
  {Mart{\'{\i}}} J.,    {Araya} E.,  2007, \aap, 465, 197

\bibitem[\protect\citeauthoryear{{Holland}}{{Holland}}{1975}]{holland75}
{Holland} J.~H.,  1975, {Adaptation in natural and artificial systems. an
  introductory analysis with applications to biology, control and artificial
  intelligence}

\bibitem[\protect\citeauthoryear{{Jarrett}, {Chester}, {Cutri}, {Schneider},
  {Skrutskie} \& {Huchra}}{{Jarrett} et~al.}{2000}]{jarrett00}
{Jarrett} T.~H.,  {Chester} T.,  {Cutri} R.,  {Schneider} S.,  {Skrutskie} M.,
    {Huchra} J.~P.,  2000, \aj, 119, 2498

\bibitem[\protect\citeauthoryear{{Joint Iras Science}}{{Joint Iras
  Science}}{1994}]{joint-iras-science94}
{Joint Iras Science} W.~G.,  1994, VizieR Online Data Catalog, 2125, 0

\bibitem[\protect\citeauthoryear{{Kenyon}, {Calvet} \& {Hartmann}}{{Kenyon}
  et~al.}{1993}]{kenyon93}
{Kenyon} S.~J.,  {Calvet} N.,    {Hartmann} L.,  1993, \apj, 414, 676

\bibitem[\protect\citeauthoryear{{Kenyon} \& {Hartmann}}{{Kenyon} \&
  {Hartmann}}{1987}]{kenyon87}
{Kenyon} S.~J.,  {Hartmann} L.,  1987, \apj, 323, 714

\bibitem[\protect\citeauthoryear{{Keto}}{{Keto}}{2002}]{keto02}
{Keto} E.,  2002, \apj, 580, 980

\bibitem[\protect\citeauthoryear{{Keto} \& {Wood}}{{Keto} \&
  {Wood}}{2006}]{keto06}
{Keto} E.,  {Wood} K.,  2006, \apj, 637, 850

\bibitem[\protect\citeauthoryear{{Keto} \& {Zhang}}{{Keto} \&
  {Zhang}}{2010}]{keto10}
{Keto} E.,  {Zhang} Q.,  2010, \mnras, pp 648--+

\bibitem[\protect\citeauthoryear{{Kim}, {Martin} \& {Hendry}}{{Kim}
  et~al.}{1994}]{kim94}
{Kim} S.,  {Martin} P.~G.,    {Hendry} P.~D.,  1994, \apj, 422, 164

\bibitem[\protect\citeauthoryear{{Lebr{\'o}n}, {Beuther}, {Schilke} \&
  {Stanke}}{{Lebr{\'o}n} et~al.}{2006}]{lebron06}
{Lebr{\'o}n} M.,  {Beuther} H.,  {Schilke} P.,    {Stanke} T.,  2006, \aap,
  448, 1037

\bibitem[\protect\citeauthoryear{{Massey} \& {Thompson}}{{Massey} \&
  {Thompson}}{1991}]{massey91}
{Massey} P.,  {Thompson} A.~B.,  1991, \aj, 101, 1408

\bibitem[\protect\citeauthoryear{{Olmi}, {Cesaroni}, {Hofner}, {Kurtz},
  {Churchwell} \& {Walmsley}}{{Olmi} et~al.}{2003}]{olmi03}
{Olmi} L.,  {Cesaroni} R.,  {Hofner} P.,  {Kurtz} S.,  {Churchwell} E.,
  {Walmsley} C.~M.,  2003, \aap, 407, 225

\bibitem[\protect\citeauthoryear{{Price}, {Egan}, {Carey}, {Mizuno} \&
  {Kuchar}}{{Price} et~al.}{2001}]{price01}
{Price} S.~D.,  {Egan} M.~P.,  {Carey} S.~J.,  {Mizuno} D.~R.,    {Kuchar}
  T.~A.,  2001, \aj, 121, 2819

\bibitem[\protect\citeauthoryear{{Qiu}, {Zhang}, {Megeath}, {Gutermuth},
  {Beuther}, {Shepherd}, {Sridharan}, {Testi} \& {De Pree}}{{Qiu}
  et~al.}{2008}]{qiu08}
{Qiu} K.,  {Zhang} Q.,  {Megeath} S.~T.,  {Gutermuth} R.~A.,  {Beuther} H.,
  {Shepherd} D.~S.,  {Sridharan} T.~K.,  {Testi} L.,    {De Pree} C.~G.,  2008,
  \apj, 685, 1005

\bibitem[\protect\citeauthoryear{{Robitaille}, {Whitney}, {Indebetouw} \&
  {Wood}}{{Robitaille} et~al.}{2007}]{robitaille07}
{Robitaille} T.~P.,  {Whitney} B.~A.,  {Indebetouw} R.,    {Wood} K.,  2007,
  \apjs, 169, 328

\bibitem[\protect\citeauthoryear{{Shakura} \& {Sunyaev}}{{Shakura} \&
  {Sunyaev}}{1973}]{shakura73}
{Shakura} N.~I.,  {Sunyaev} R.~A.,  1973, \aap, 24, 337

\bibitem[\protect\citeauthoryear{{Shepherd}, {Yu}, {Bally} \&
  {Testi}}{{Shepherd} et~al.}{2000}]{shepherd00}
{Shepherd} D.~S.,  {Yu} K.~C.,  {Bally} J.,    {Testi} L.,  2000, \apj, 535,
  833

\bibitem[\protect\citeauthoryear{{Shu}, {Najita}, {Ostriker}, {Wilkin}, {Ruden}
  \& {Lizano}}{{Shu} et~al.}{1994}]{shu94}
{Shu} F.,  {Najita} J.,  {Ostriker} E.,  {Wilkin} F.,  {Ruden} S.,    {Lizano}
  S.,  1994, \apj, 429, 781

\bibitem[\protect\citeauthoryear{{Shu}, {Adams} \& {Lizano}}{{Shu}
  et~al.}{1987}]{shu87}
{Shu} F.~H.,  {Adams} F.~C.,    {Lizano} S.,  1987, \araa, 25, 23

\bibitem[\protect\citeauthoryear{{Sridharan}, {Beuther}, {Saito}, {Wyrowski} \&
  {Schilke}}{{Sridharan} et~al.}{2005}]{sridharan05}
{Sridharan} T.~K.,  {Beuther} H.,  {Saito} M.,  {Wyrowski} F.,    {Schilke} P.,
   2005, \apjl, 634, L57

\bibitem[\protect\citeauthoryear{{Su}, {Liu}, {Chen}, {Zhang} \&
  {Cesaroni}}{{Su} et~al.}{2007}]{su07}
{Su} Y.,  {Liu} S.,  {Chen} H.,  {Zhang} Q.,    {Cesaroni} R.,  2007, \apj,
  671, 571

\bibitem[\protect\citeauthoryear{{Terebey}, {Shu} \& {Cassen}}{{Terebey}
  et~al.}{1984}]{terebey84}
{Terebey} S.,  {Shu} F.~H.,    {Cassen} P.,  1984, \apj, 286, 529

\bibitem[\protect\citeauthoryear{{Toomre}}{{Toomre}}{1964}]{toomre64}
{Toomre} A.,  1964, \apj, 139, 1217

\bibitem[\protect\citeauthoryear{{Ulrich}}{{Ulrich}}{1976}]{ulrich76}
{Ulrich} R.~K.,  1976, \apj, 210, 377

\bibitem[\protect\citeauthoryear{{van der Tak}, {van Dishoeck}, {Evans} II \&
  {Blake}}{{van der Tak} et~al.}{2000}]{van-der-tak00}
{van der Tak} F.~F.~S.,  {van Dishoeck} E.~F.,  {Evans} II N.~J.,    {Blake}
  G.~A.,  2000, \apj, 537, 283

\bibitem[\protect\citeauthoryear{{Weingartner} \& {Draine}}{{Weingartner} \&
  {Draine}}{2001}]{weingartner01}
{Weingartner} J.~C.,  {Draine} B.~T.,  2001, \apj, 548, 296

\bibitem[\protect\citeauthoryear{{Whitney}, {Indebetouw}, {Bjorkman} \&
  {Wood}}{{Whitney} et~al.}{2004}]{whitney04c}
{Whitney} B.~A.,  {Indebetouw} R.,  {Bjorkman} J.~E.,    {Wood} K.,  2004,
  \apj, 617, 1177

\bibitem[\protect\citeauthoryear{{Whitney}, {Wood}, {Bjorkman} \&
  {Wolff}}{{Whitney} et~al.}{2003}]{whitney03}
{Whitney} B.~A.,  {Wood} K.,  {Bjorkman} J.~E.,    {Wolff} M.~J.,  2003, \apj,
  591, 1049

\bibitem[\protect\citeauthoryear{{Wilking}, {Blackwell} \& {Mundy}}{{Wilking}
  et~al.}{1990}]{wilking90}
{Wilking} B.~A.,  {Blackwell} J.~H.,    {Mundy} L.~G.,  1990, \aj, 100, 758

\bibitem[\protect\citeauthoryear{{Williams}, {Fuller} \&
  {Sridharan}}{{Williams} et~al.}{2005}]{williams05}
{Williams} S.~J.,  {Fuller} G.~A.,    {Sridharan} T.~K.,  2005, \aap, 434, 257

\bibitem[\protect\citeauthoryear{{Wood}, {Lada}, {Bjorkman}, {Kenyon},
  {Whitney} \& {Wolff}}{{Wood} et~al.}{2002}]{wood02b}
{Wood} K.,  {Lada} C.~J.,  {Bjorkman} J.~E.,  {Kenyon} S.~J.,  {Whitney} B.,
  {Wolff} M.~J.,  2002, \apj, 567, 1183

\bibitem[\protect\citeauthoryear{{Wood}, {Wolff}, {Bjorkman} \&
  {Whitney}}{{Wood} et~al.}{2002}]{wood02a}
{Wood} K.,  {Wolff} M.~J.,  {Bjorkman} J.~E.,    {Whitney} B.,  2002, \apj,
  564, 887

\bibitem[\protect\citeauthoryear{{Yorke}}{{Yorke}}{2002}]{yorke02}
{Yorke} H.~W.,  2002, in {Crowther} P.,  ed., Hot Star Workshop III: The
  Earliest Phases of Massive Star Birth Vol.~267 of Astronomical Society of the
  Pacific Conference Series, {Theory of Formation of Massive Stars via
  Accretion}.
pp 165--+

\bibitem[\protect\citeauthoryear{{Yusef-Zadeh}, {Morris} \&
  {White}}{{Yusef-Zadeh} et~al.}{1984}]{yusef-zadeh84}
{Yusef-Zadeh} F.,  {Morris} M.,    {White} R.~L.,  1984, \apj, 278, 186

\bibitem[\protect\citeauthoryear{{Zhang}, {Hunter} \& {Sridharan}}{{Zhang}
  et~al.}{1998}]{zhang98}
{Zhang} Q.,  {Hunter} T.~R.,    {Sridharan} T.~K.,  1998, \apjl, 505, L151

\end{thebibliography}
\end{document}